
\input amstex

\NoBlackBoxes
\documentstyle{amsppt}
\document

\centerline {\bf Representations of affine Lie algebras,}
\centerline {\bf elliptic $r$-matrix systems,}
\centerline{\bf and special functions}
\vskip .15 in
\centerline{\bf Pavel I. Etingof}
\vskip .1in
\centerline{Department of Mathematics}
\centerline{Yale University}
\centerline{Box 2155, Yale Station}
\centerline{New Haven, CT 06515}
\centerline{e-mail etingof\@ pascal.math.yale.edu}
\vskip .1in

\centerline{Submitted to Comm. Math. Phys. on JANUARY 28, 1993}
\vskip .1in

\heading
{\bf Abstract}
\endheading

The author considers an elliptic analogue of the
Knizhnik-Zamolodchikov equations --
the consistent system of linear differential
equations arising from the elliptic solution of the classical
Yang-Baxter equation for the Lie algebra ${\frak sl}_N$.

The solutions
of this system are interpreted as traces of
products of intertwining operators between certain representations
of the affine Lie algebra $\widehat{\frak sl}_N$.

A new differential equation for such traces characterizing their behavior
under the variation of the modulus of the underlying elliptic curve
is deduced. This equation is consistent with the original system.

It is shown that the system extended by the new equation is modular
invariant, and the corresponding monodromy representations of the
modular group are defined.

Some elementary examples in which the
system can be solved explicitly (in terms of elliptic and modular
functions) are considered. Another example leads to a special case of
Heun's equation -- a second order Fuchsian equation with four singular
points.

 The monodromy of the system is explicitly
computed  with the help of the trace interpretation of solutions.
Projective representations of the
braid group of the torus and representations of the double affine
Hecke algebra are obtained.
\vskip .2in

\heading
{\bf Introduction}
\endheading

In 1984 Knizhnik and Zamolodchikov \cite{KZ} studied matrix elements
of
products
of intertwining operators between representations of the affinization
$\hat{\frak g}$ of a finite dimensional simple complex Lie
algebra $\frak g$ at level $k$.
These matrix elements are analytic functions of several complex
variables, and it was found that they satisfy a certain remarkable
system of linear differential equations which is now called the
Knizhnik-Zamolodchikov (KZ) system:
$$
(k+h^{\vee})\frac{\partial \Psi}{\partial z_i}=\sum_{j=1,j\ne
i}^n\frac{\Omega_{ij}}{z_i-z_j}\Psi.\tag 1
$$
Here $\Psi(z_1,...,z_n)$ is a function of $n$ complex variables
with values in the product $W=V_1\otimes V_2\otimes \dots\otimes V_n$
of $n$ representations of $\frak g$, $h^{\vee}$
is the dual Coxeter number of $\frak g$, and
$$
\gather
\Omega_{ij}=\sum_{p}(x_p)_i(x_p)_j, \\
(x_p)_i=\text{Id}_1\otimes\dots\otimes
\text{Id}_{i-1}\otimes x_p\otimes
\text{Id}_{i+1}\otimes\dots\otimes\text{Id}
_n
\in \text{End}(W),\tag 2
\endgather
$$
where the summation is over
 an orthonormal base $\lbrace x_p\rbrace$ of $\frak g$ with respect to
the invariant form, and $\text{Id}_j$ denotes the identity operator in $V_j$.

 Solutions of the KZ equations
are very interesting special functions which generalize the Gauss
hypergeometric function. By now an explicit integral representations
of these functions has been found \cite{M},\cite{SV},
and their monodromy has been completely computed
\cite{Koh},\cite{Dr},\cite{TK} (for ${\frak g}={\frak
sl}_2$),\cite{V}.
It turned out that the monodromy of the KZ system
expresses in terms of the quantum $R$-matrix -- the
quasitriangular structure of the corresponding quantum group
$U_q(\frak g)$, where $q=\exp\bigl(\frac{2\pi\text{i}}{k+h^{\vee}}\bigr)$.

Cherednik \cite{Ch1} considered a general consistent system of
differential equations of the ``factorized'' form
$$
\kappa\frac{\partial \Psi}{\partial z_i}=\sum_{j=1,j\ne
i}^nr_{ij}(z_i-z_j)\Psi,\tag 3
$$
where $r(u)$ is a meromorphic function with values in ${\frak
g}\otimes {\frak g}$, and $r_{ij}(u)$ denotes the action of $r(u)$ in $W$:
the first factor acts in $V_i$ and the second one in $V_j$, and the
{\it unitarity} property $r_{ij}(u)=-r_{ji}(-u)$ is assumed
(this condition is equivalent to the
invariance of solutions of system (3) under the simultaneous translation of all
variables $z_j$ by the same constant).
It was observed in \cite{Ch1} that system
(3) is consistent if and only if the function $r(u)$ is a {\it
classical $r$-matrix}, i.e. if it satisfies the classical Yang-Baxter
equation:
$$
[r_{ij}(z_i-z_j),r_{ik}(z_j-z_k)]+
[r_{ij}(z_j-z_k),r_{jk}(z_k-z_i)]+
[r_{ik}(z_k-z_i),r_{jk}(z_i-z_j)]=0.\tag 4
$$
Therefore, a consistent system (3) is called a local $r$-matrix
system.

 Clearly,
the KZ system is a special case of a local $r$-matrix system,
for a simplest $r$-matrix $r(u)=\Omega/u$,
$\Omega=\sum_px_p\otimes x_p$,
(the summation is over an orthonormal base $\lbrace x_p\rbrace$
of $\frak g$). This gives
rise to a question: what other $r$-matrix systems are possible?
This question was essentially answered by Belavin and Drinfeld in 1982
\cite{BeDr}. They classified all solutions of (3) satisfying
the nondegeneracy condition: $r(u)$ is invertible as a map
${\frak g^*}\to {\frak g}$ for at least one complex number $u$.
This classification states that all such solutions are unitary and,
in terms of dependence on $u$, there
are only three types of functions $r(u)$: rational, trigonometric,
and elliptic. For instance, the function $r(u)=\Omega/u$ which is
involved in the KZ system is a rational
$r$-matrix.

Trigonometric an elliptic nondegenerate unitary solutions of the
classical Yang-Baxter equation are completely classified \cite{BeDr}.
On the contrary, a satisfactory classification of rational solutions
is unknown.

The Belavin-Drinfeld classification suggests a two-step generalization
of the KZ system: KZ equations (rational $r$-matrix equations)
-- trigonometric $r$-matrix
equations -- elliptic $r$-matrix equations. One should expect that these
local $r$-matrix systems should have remarkable properties
and provide new interesting special functions as their solutions.

Cherednik \cite{Ch2} found an interpretation of solutions of
nondegenerate unitary
$r$-matrix equations in terms of representation theory of affine Lie
algebras. It was proved in \cite{Ch2} that the general solution is the
coinvariant (or the so-called $\tau$-function) of
the Lie algebra of $\frak g$-valued rational functions
on a rational or elliptic curve with singularities at designated
points $z_1,...,z_n$ in a certain representation of this algebra.
This interpretation has found many applications.

The properties of the trigonometric $r$-matrix equations
are now fairly well understood. Let ${\frak g}={\frak n}^+\oplus{\frak
h}\oplus{\frak n}^-$ be a Cartan decomposition of $\frak g$, and let
$\hat\Omega^+$, $\hat\Omega^-$, $\hat\Omega^0$ be the orthogonal
projections of $\Omega$ to the subspaces ${\frak n}^+\otimes{\frak
n}^-$, ${\frak n}^-\otimes{\frak n}^+$, and ${\frak h}\otimes{\frak
h}$, respectively . Then the
simplest trigonometric ${\frak g}\otimes {\frak g}$-valued
$r$-matrix has the form $r(u)=\frac{\Omega^+e^u+\Omega^-}{u-1}$, where
$\Omega^{\pm}=\hat\Omega^{\pm}+\frac{1}{2}\hat\Omega_0$.
It turns out that
there exists a transformation of coordinates which maps the
$n+1$-point KZ equations to the $n$-point trigonometric $r$-matrix
equations with the above $r(u)$ (which are called the trigonometric KZ
equations), which allows one to carry over to
this case all the results about the KZ equations, including the
integral formulas.

Trigonometric $r$-matrix equations with more general $r$-matrices
were studied by Cherednik \cite{Ch2}, who
gave an explicit integral formula for the general solution.

 Nondegenerate unitary
elliptic $r$-matrices exist only for ${\frak g}={\frak
sl}_N$. They were found in the case $N=2$ by Sklyanin and in the
general case by Belavin\cite{Be}
(note that the $N=2$ elliptic $r$-matrix is the quasiclassical limit of
 Baxter`s quantum $R$-matrix which arises in statistical mechanics):
$$
\rho(z|\tau)=\Omega\zeta(z)+\sum_{0\le m,n\le N-1,m^2+n^2>0}(1\otimes
\beta^n\gamma^{-m})(\Omega)\bigl(\zeta(z-\frac{m+n\tau}{N}|\tau)-\zeta(\frac{m+n\tau}{N}|\tau)
\bigr),\tag 5
$$
 where $\beta,\gamma$ are two commuting inner automorphisms of ${\frak sl}_N$
of order $N$ with no common invariant vectors, $\tau$ lies in the
upper half of the complex plane, and $\zeta(z|\tau)$ is the
Weierstrass $\zeta$-function.
 The properties of the corresponding
$r$-matrix equations are not very well understood. For example,
integral formulas (or any other explicit representations) for
solutions are unknown.

It has been anticipated that some progress in the elliptic case can be
achieved by using the intertwining (vertex) operator language which was
originally used by Knizhnik and Zamolodchikov\cite{KZ}.
Frenkel and Reshetikhin \cite{FR} conjectured that if one takes traces
of products of intertwiners rather than matrix elements, one should be
able to obtain solutions to the elliptic $r$-matrix equations.
The same idea occurs in the paper of Bernard \cite{Ber} who studied
expectation values for
the Wess-Zumino-Witten model on an elliptic curve and was led to consider
trace expressions of a similar sort. Bernard deduced some differential
relations for traces, but they were not a closed system of
differential equations since the two commuting inner automorphisms
did not enter the game. The idea to consider traces is also suggested
by the aforementioned theory developed in \cite{Ch2} which states that
solutions of an elliptic $r$-matrix system should express in terms of
the $\tau$-function for the corresponding elliptic curve, which is
basically a trace expression involving vertex operators.

This paper is devoted to making this idea work
(at the price of a few modifications). More precisely,
we represent solutions of the elliptic $r$-matrix equations
in the form
$$
F(z_1,...,z_n|q)=\text{Tr}\mid
_{M_{\lambda,k}}(\Phi^1(z_1)\dots\Phi^n(z_n)Bq^{-\partial}),\tag
6
$$
where $M_{\lambda,k}$ is a Verma module over (a twisted version of)
the affine algebra $\widehat{\frak sl}_N$, $\Phi(z_i):\ M_{\lambda_i,k}\to
\hat M_{\lambda_{i-1},k}\otimes V_i$ are intertwiners for $\widehat{\frak
sl}_N$ (the action of $\widehat{\frak sl}_N$ on $V_i$: $(a\otimes
t^m)v=z^mav$, $a\in\frak g$, $v\in V_i$; the hat denotes a completion
of the Verma module),
$q\in\Bbb C$, $|q|<1$,
$\partial$ is the grading operator in $M_{\lambda,k}$, and $B$ is the
map of Verma modules induced by an outer automorphism of
$\widehat{\frak sl}_N$ of order $N$ (rotation of the affine Dynkin
diagram, which is a regular $N$-gon, through the angle of $2\pi/N$).
This representation helps to compute the monodromy of the elliptic
$r$-matrix equations. It is still unclear how to deduce
integral formulas for solutions similar to those existing for the
rational and trigonometric KZ equations.

{\bf Remark. } The recent paper \cite{CFW} studies generalized hypergeometric
functions on the torus -- integrals over twisted cycles of products of
powers of elliptic functions. These functions satisfy certain linear
differential equations with elliptic coefficients, but it is not clear
how these equations are related to the elliptic $r$-matrix equations.

In Section 1 we introduce a realization of an affine Lie
algebra twisted by an inner automorphism
and twisted versions of Verma modules and evaluation modules.
This twisting is necessary to eventually produce solutions of the
elliptic $r$-matrix equations.

In Section 2 we define twisted intertwiners and deduce a differential
equation for them. Then we define twisted correlation functions
and show that they satisfy a twisted trigonometric KZ system. This system,
however, can be reduced to the usual KZ system by a simple
transformation.

In Section 3 we study trace expressions of the form (6) and prove that
they satisfy an elliptic $r$-matrix system of differential equations
in the variables $\log z_j$. (We call this system the elliptic KZ
equations) We also deduce one more differential
equation which expresses the derivative $\frac{\partial F}{\partial q}$ in
terms of $F$. Thus we get a consistent system of $n+1$ differential
equations -- the extended elliptic KZ system.

In Section 4 we show that the elliptic KZ equations are modular
invariant -- they are preserved under the action of the congruence
subgroup $\Gamma(N)$ of the modular group. The $n+1$-th equation
involving the derivative by $q$ is ``almost'' invariant -- it
undergoes a very minor modification under the action of an element of
$\Gamma(N)$. This implies that the fundamental solution of the
extended elliptic KZ system changes under the action of the modular
group according to a certain representation of this group
. In other
words, the functions $F(z_1,...,z_n|q)$ yield nontrivial examples of
vector-valued automorphic (modular) forms.

In Section 5 we consider a few simple special cases of the
extended elliptic KZ system for ${\frak sl}_2$
in which it can be solved in quadratures.
In this case, solutions express in terms of elliptic and modular functions.
Example 4 treats a slightly more complicated case in which the
elliptic KZ system reduces to a special (but non-integrable) case of
the so-called Heun's equation -- the general second order linear
differential equation on
$\Bbb CP^1$ with four regular singular points.

In section 6 we compute the monodromy of the elliptic KZ equations.
According to the results of \cite{Ch1}, this monodromy yields
a representation of the generalized braid group of the torus.
We compute this representation and show that local monodromies (around
the loci $z_i=z_j$) are the same as for the usual KZ equations and can be
described in terms of the quantum $R$-matrix -- a known result
from the theory of $r$-matrix systems
\cite{Ch1}. Global monodromies (around the $\tau$-cycle
of the elliptic curve) are described in terms
of ordered products of $R$-matrices, of the type
occuring in the quantum KZ equation (see \cite{FR}).
In the special case when for all $j$ $V_j$ is the $N$-dimensional
vector representation of ${\frak sl}_N$ -- the monodromy
representation is a representation of
the double affine Hecke algebra recently introduced by
Cherednik\cite{Ch3}. As an aside,
the examination of monodromy helps to prove that
if $\kappa=\frac{1}{NM}$ where $M$ is an integer then the elliptic
KZ equations are integrable in elliptic functions.

A generalization of the results of this paper to the case of a quantum
affine algebra will be described in a forthcoming paper.

\heading
{\bf Acknowledgements}
\endheading

I would like to thank my adviser I.Frenkel for stimulating my
work by systematic inspiring discussions. I am grateful to
I.Cherednik, W.Feit, H.Garland,
D.Kazhdan, and A.Varchenko for many useful suggestions,
and to A.Kirillov Jr.
for sharing with me the results of his previous work on the subject
and for illuminating remarks.

\vskip .1in

\heading
{\bf 1. A twisted realization of affine Lie algebras.} \endheading

Let $\frak g$ be a finite dimensional simple Lie algebra over
$\Bbb C$ of rank $r$.
Denote by $<,>$ the standard invariant form on $\frak g$ with respect to which
the longest root has length $\sqrt{2}$.

Let $\frak h$ denote a Cartan subalgebra of $\frak g$. The form $<,>$
defines a natural identification ${\frak h^*}\to{\frak h}$:
$\lambda\mapsto h_{\lambda}$ for $\lambda\in{\frak h}^*$. We will
use the notation $<,>$ for the inner product in both $\frak h$ and
$\frak h^*$.

Let $\Delta^+$ be the set of positive roots of $\frak g$.
For $\alpha\in\Delta^+$, let $e_{\alpha}$, $f_{\alpha}$, $H_{\alpha}$ be the
standard basis of the ${\frak sl}_2$-subalgebra in $\frak g$ associated with
$\alpha$: $[H_{\alpha},e_{\alpha}]=2e_{\alpha},\
[H_{\alpha},f_{\alpha}]=-2f_{\alpha},\
[e_{\alpha},f_{\alpha}]=H_{\alpha}$.
Let $|\alpha|$ denote the number of summands in the decomposition of
$\alpha$ in the sum of simple positive roots.

Let $N$ be the dual Coxeter
number of $\frak g$. Let
$\rho=\frac{1}{2}\sum_{\alpha\in\Delta^+}\alpha$.
The elements $\rho$ and $h_{\rho}$ satisfy the relations
$\rho(h_{\alpha})=\alpha(h_{\rho})=|\alpha|$
(note that in general $H_{\alpha}\ne h_{\alpha}$).

Let $\gamma$ be an inner automorphism of $\frak g$:
$\gamma(a)=\text{Ad} C(a)$, $a\in\frak g$,
where $C=e^{\frac{2\pi \text{i}lh_{\rho}}{N}}$, $l\in\Bbb Z$, $1\le l<N$,
and $l,N$ are
coprime\footnote{Note that throughout the paper we denote the complex
number $\text{i}=\sqrt{-1}$ by a roman ``i'', to distinguish it from
the subscript $i$, which is italic.}. This automorphism is of order
$N$.

The action of $\gamma$ on root vectors is as follows:
$\gamma(e_{\alpha})=\varepsilon^{|\alpha|}e_{\alpha}$
, $\gamma(f_{\alpha})=\varepsilon^{-|\alpha|}f_{\alpha}$
, where $\varepsilon=e^{2\pi \text{i}l/N}$ is a primitive $N$-th root of unity.

Let $x_j$, $1\le j\le r$,
be an orthonormal basis of $\frak h$ with respect to the standard
invariant form.

Let $\hat{\frak g}={\frak g}\otimes \Bbb C[t,t^{-1}]\oplus \Bbb Cc$
be the affine Lie algebra associated with $\frak g$. The commutation relations
in this algebra are
$$
[a(t)+\lambda c,b(t)+\mu c]=[a(t),b(t)]+\frac{1}{2\pi \text{i}}
\oint_{|t|=1}<a^{\prime}(t)b(t)>t^{-1}dt\cdot c\tag 1.1
$$
for any two $\frak g$-valued Laurent polynomials $a(t)$, $b(t)$, and
complex numbers $\lambda$, $\mu$ \cite{Ka}.
The elements $e_{\alpha}\otimes t^m$, $f_{\alpha}\otimes t^m$,
$x_i\otimes t^m$, $c$, for $m\in \Bbb Z$, $\alpha\in\Delta^+$, form a basis of
$\hat {\frak g}$.

Define the subalgebra $\hat{\frak g}_{\gamma}$ of $\hat{\frak g}$ consisting
of all expressions $a(t)+\lambda c$ with the property
$a(\varepsilon t)=\gamma(a(t))$.

\proclaim {Lemma 1.1} (see \cite{PS},p.36) The Lie algebras
$\hat{\frak g}_{\gamma}$ and
$\hat{\frak g}$ are isomorphic.
\endproclaim

\demo{Proof}
The elements
$e_{\alpha}\otimes t^{|\alpha|+mN}$, $f_{\alpha}\otimes t^{-|\alpha|+mN}$,
$x_i\otimes t^{mN}$, $c$, for $m\in \Bbb Z$, $\alpha\in\Delta^+$, form a basis
of $\hat {\frak g}_{\gamma}$. Define a map $\phi:\hat{\frak
g}_{\gamma}\to\hat{\frak g}$ by:
$$
\gather
\phi(e_{\alpha}\otimes t^{|\alpha|+mN})=e_{\alpha}\otimes t^m,\\
\phi(f_{\alpha}\otimes t^{-|\alpha|+mN})=f_{\alpha}\otimes t^m,\\
\phi(x_j\otimes t^{mN})=x_j\otimes t^m, m\ne 0,\\
\phi(x_i)=x_i-\frac{1}{N}\rho(x_i)c,\ \phi(c)=\frac{c}{N}.\tag 1.2\endgather
$$

It is easy to check that $\phi$ is an isomorphism of Lie algebras.
\enddemo

Thus, the twisting of $\hat{\frak g}$ by $\gamma$ does not give us a new Lie
algebra. However, the twisted realization $\hat{\frak g}_{\gamma}$
of the affinization of $\frak g$ will be very convenient in further
considerations.

Let us translate some well known results about representations of
$\hat{\frak g}$ into the "twisted" language.

First of all, define the polarization of  $\hat{\frak g}_{\gamma}$:
$\hat{\frak g}_{\gamma}=\hat{\frak g}_{\gamma}^+\oplus\hat{\frak g}_{\gamma}^-
\oplus {\frak h}\oplus \Bbb Cc$. Here $\hat{\frak g}_{\gamma}^+$ is the set of
polynomials $a(t)$ vanishing at $0$, and $\hat{\frak g}_{\gamma}^-$ is the set
of polynomials $a(t)$ vanishing at infinity.

Next, define Verma modules over $\hat{\frak g}_{\gamma}$. This is done exactly
in the same way as for the untwisted affine algebra.
Let $\lambda\in \frak h^*$ be a weight, and let $k$ be a complex number.
Define $X_{\lambda,k}$ to be a one dimensional module over
$\hat{\frak g}_{\gamma}^+\oplus {\frak h}\oplus \Bbb Cc$ spanned by a vector
$v$ such that $\hat{\frak g}_{\gamma}^+$ annihilates $v$, and
$cv=kv$, $hv=\lambda(h)v$, $h\in \frak h$. Define the Verma module
$$
M_{\lambda,k}=\text{Ind}^{\hat{\frak g}_{\gamma}}_{\hat{\frak g}_{\gamma}^+
\oplus {\frak h}\oplus \Bbb Cc}X_{\lambda,k}.\tag 1.3
$$

Now define evaluation representations.
Let $V$ be a highest weight module over $\frak g$.
It will be convenient to assume that $V$ is a quotient of a Verma module.
Define the operator $C\in\text{End}(V)$ by the conditions:
$Caw=\gamma(a)Cw$ for any $w\in V$, $a\in\frak g$, and
$Cw_0=w_0$, where $w_0$ is the highest weight vector of $V$.

Let $V(z)$ denote the space of $V$-valued Laurent polynomials in $z$, and let
$V_C(z)$ be the space of those polynomials which satisfy the equivariance
condition $w(\varepsilon z)=Cw(z)$.

The natural (pointwise) action of $\hat{\frak g}$ on $V(z)$ restricts
to an action of $\hat{\frak g}_{\gamma}$ on $V_C(z)$.
For this twisted action we have an analogue of Lemma 1.1.

\proclaim{Lemma 1.2} The isomorphism $\phi$ transforms the
$\hat{\frak g}_{\gamma}$-module $V_C(z)$ into a $\hat{\frak g}$-module,
isomorphic to $V(z)$.
\endproclaim

Let us introduce the twisted version of currents.
Set
$$
\gather
J_{e_{\alpha}}(z)=\sum_{m\in \Bbb Z}e_{\alpha}\otimes t^{|\alpha|+mN}\cdot
z^{-|\alpha|-mN-1},\\
J_{f_{\alpha}}(z)=\sum_{m\in \Bbb Z}f_{\alpha}\otimes t^{-|\alpha|+mN}\cdot
z^{|\alpha|-mN-1},\\
J_h(z)=\sum_{m\in \Bbb Z}h\otimes t^{mN}\cdot z^{-mN-1},\ h\in{\frak h}.\tag
1.4\endgather
$$
Thus by linearity we have defined $J_a(z)$ for any $a\in {\frak g}$.

Define the polarization of currents:
$$
\gather
J^+_{e_{\alpha}}(z)=\sum_{m<0}e_{\alpha}\otimes t^{|\alpha|+mN}\cdot z^{-
|\alpha|-mN-1},\\
J^+_{f_{\alpha}}(z)=\sum_{m\le 0}f_{\alpha}\otimes t^{-|\alpha|+mN}\cdot
z^{|\alpha|-mN-1},\\
J^+_h(z)=\frac{1}{2}h\otimes 1\cdot z^{-1}+\sum_{m<0}h\otimes t^{mN}\cdot z^{-
mN-1},\ h\in{\frak h}.\tag 1.5\endgather
$$
This defines $J_a^+(z)$ for all $a\in\frak g$. Now set
$$
J_a^-(z)=J_a^+(z)-J_a(z).\tag 1.6
$$

Note that this polarization is not quite
the same as the standard polarization of currents for the untwisted
$\hat{\frak g}$\cite{Ka}, i.e. the isomorphism $\phi$ does not match up these
two polarizations.

The Lie algebra $\hat{\frak g}_{\gamma}$ has a natural $\Bbb Z$-grading
by powers of $t$. Thus every Verma module $M_{\lambda,k}$ is natuarally
$\Bbb Z$-graded: the degree of the highest weight vector $v$ is zero,
and for every homogeneous vector $w$ $\text{deg}(a\otimes t^{-
m}w)=\text{deg}(w)-m$.
Let $d\in \text{End}(M_{\lambda,k})$ be the grading operator:
if $w$ is a homogeneous vector then $dw=\text{deg}(w)w$. The operator $d$
satisfies
the commutation relations
$[d,a(t)]=ta^{\prime}(t)$, $[d,c]=0$.

Let us find an expression for $d$ in terms of elements of $\hat {\frak g}$ --
a twisted version of the Suguwara construction. We assume that $k\ne -1$.

\proclaim{Proposition 1.3}
$$
d=-\frac{1}{k+1}\sum_{m\in\Bbb Z}\left(\sum_{\alpha\in\Delta^+}
:e_{\alpha}\otimes t^{|\alpha|+mN}f_{\alpha}\otimes t^{-|\alpha|-mN}:
+\frac{1}{2}\sum_{j=1}^{r}:x_j\otimes t^{mN}x_j\otimes t^{-mN}:\right)+
\frac{<\lambda,\lambda>}{2(k+1)},\tag 1.7
$$
where :: is the standard normal ordering:
$$\gather
:e_{\alpha}\otimes t^nf_{\alpha}\otimes t^{-n}:=\cases
 e_{\alpha}\otimes t^nf_{\alpha}\otimes t^{-n}
 & n<0\\ f_{\alpha}\otimes t^{-n}e_{\alpha}\otimes t^n
 & n>0\endcases
\\
:h\otimes t^nh\otimes t^{-n}:=\cases h\otimes t^nh\otimes t^{-n}&n\le 0\\
h\otimes t^{-n}h\otimes t^n &n>0 \endcases,\ h\in{\frak h}.
\tag 1.8
\endgather
$$
\endproclaim

\demo{Proof} Let $\Cal M_{\Lambda,K}$ be the Verma module over $\hat{\frak g}$
with highest weight $\Lambda$ and central charge $K$. Lemma 1.1 implies that
the isomorphism $\phi$ transforms the module $M_{\lambda,k}$ over
$\hat{\frak g}_{\gamma}$ into the module $\Cal M_{\Lambda,K}$ over $\hat{\frak
g}$, with $K=Nk$, $\Lambda=\lambda+k\rho$. Let $D$ be the grading operator
in $\Cal M_{\Lambda,K}$ which is associated with the grading of $\hat{\frak
g}$ by powers of $t$. Then, according to Lemma 1.1, $\phi(d)=ND+h_{\rho}$.
Therefore, $$d=\phi^{-1}(ND+h_{\rho})=N\phi^{-1}(D)+h_{\rho}+<\rho,\rho>k\tag
1.9$$

For the standard affine algebra $\hat{\frak g}$, the operator $D$ is given by
the Suguwara formula:
$$
D=-\frac{1}{K+N}\sum_{m\in\Bbb Z}\left(\sum_{\alpha\in\Delta^+}
:e_{\alpha}\otimes t^mf_{\alpha}\otimes t^{-m}:
+\frac{1}{2}\sum_{j=1}^{r}:x_j\otimes t^mx_j\otimes t^{-m}:\right)+
\frac{<\lambda,\lambda+2\rho>}{2(K+N)},\tag 1.10
$$
where the normal ordering is defined by (1.8) and
$$
:e_{\alpha}f_{\alpha}:=
\frac{1}{2}(e_{\alpha}f_{\alpha}+f_{\alpha}e_{\alpha}).\tag 1.11
$$

Substituting (1.10) into (1.9), after some algebra we get
(1.7).
\enddemo

Let us now extend the Lie algebra $\hat{\frak g}_{\gamma}$
by adding a new element $\partial$ satisfying the relations
$[\partial, a(t)]=ta^{\prime}(t)$, $[\partial,c]=0$.
Denote the obtained Lie algebra by $\tilde{\frak g}_{\gamma}$.
Then the action of $\hat{\frak g}_{\gamma}$
in every Verma module $M_{\lambda,k}$ extends to an action of
$\tilde{\frak g}_{\gamma}$: one sets
$\partial=d-\frac{<\lambda,\lambda>}{2(k+1)}$ (to get rid of the free
term in (1.7)). The action of $\partial$ can
also be defined in evaluation representations $V_C(z)$:
$\partial=z\frac{d}{dz}$. Thus $V_C(z)$ becomes a
$\tilde{\frak g}_{\gamma}$-module.
\vskip .1in

\heading
{\bf 2. Twisted intertwiners and Knizhnik-Zamolodchikov equations}
\endheading

We will be interested in $\tilde{\frak g}_{\gamma}$
intertwining operators $\Phi(z):M_{\lambda,k}\to M_{\nu,k}\hat\otimes
z^{\Delta}V_C(z)$, where the highest weight of $V$ is $\mu$,
$\hat\otimes$ denotes the completed tensor product, and $\Delta$ is a
complex number.

Lemma 1.1 and the results of the untwisted theory
imply the following statement:

\proclaim{Proposition 2.1} Operators $\Phi$ are in one-to-one correspondence
with vectors in $V$ of weight $\lambda-\nu$.
This correspondence is defined by the action of $\Phi$ at the vacuum level.
\endproclaim

Let $z_0$ be a nonzero complex number. Evaluation of the operator
$\Phi(z)$ at the point $z_0$ yields an operator $\Phi(z_0):M_{\lambda,k}\to
\hat M_{\nu,k}\otimes V$, where $\hat M$ denotes the completion of $M$
with respect to the grading.

 From now on the notation $\Phi(z)$ will mean
the operator
$\Phi$ evaluated at the point $z\in\Bbb C^*$. This will give us an opportunity
to consider the operator $\Phi(z)$ as an analytic function of $z$. This
analytic function will be multivalued: $\Phi(z)=z^{\Delta}\Phi^0(z)$, where
$\Phi^0$ is a single-valued function in $\Bbb C^*$, and
$\Delta=\frac{<\nu,\nu>-<\lambda,\lambda>}{2(k+1)}$.

Let $u$ belong to the restricted dual module $V^*$. Introduce the notation
\linebreak $\Phi_u(z)=(1\otimes u)(\Phi(z))$. $\Phi_u(z)$ is an
operator: $M_{\lambda,k}\to
\hat M_{\nu,k}$.

The intertwining property for $\Phi(z)$
can be written in the form
$$
[a\otimes t^m,\Phi_u(z)]=z^m\Phi_{au}(z).\tag 2.1
$$

It is convenient to write the intertwining relation in terms of currents.

\proclaim{Lemma 2.2}
$$
\gather
[J_h^{\pm}(\zeta),\Phi_u(z)]=\frac{1}{2\zeta}\frac{\zeta^N+z^N}{z^N-
\zeta^N}\Phi_{hu}(z),\ h\in{\frak h};\\
[J_{e_{\alpha}}^{\pm}(\zeta),\Phi_u(z)]=\frac{\zeta^{N-1-|\alpha|}z^{|\alpha|}
}{z^N-\zeta^N}\Phi_{e_{\alpha}u}(z),\ \alpha\in\Delta_+;\\
[J_{f_{\alpha}}^{\pm}(\zeta),\Phi_u(z)]=\frac{\zeta^{|\alpha|-1}z^{N-|\alpha|}
}{z^N-\zeta^N}\Phi_{f_{\alpha}u}(z),\ \alpha\in\Delta_+.\tag 2.2
\endgather
$$
\endproclaim

The identities marked with $+$ make sense if $|z|>|\zeta|$, and those
marked with $-$ make sense if $|z|<|\zeta|$.

Now we are ready to write down the twisted version of the operator Knizhnik-
Zamolodchikov (KZ) equations.

\proclaim{Theorem 2.3} The operator function $\Phi_u(z)$ satisfies the
differential equation
$$
\gather
(k+1)\frac{d}{dz}\Phi_u(z)=\sum_{\alpha\in\Delta^+}
(J_{e_{\alpha}}^+(z)\Phi_{f_{\alpha}u}(z)-\Phi_{f_{\alpha}u}(z)
 J_{e_{\alpha}}^-(z))+\\
\sum_{\alpha\in\Delta^+}
(J_{f_{\alpha}}^+(z)\Phi_{e_{\alpha}u}(z)-\Phi_{e_{\alpha}u}(z)
 J_{f_{\alpha}}^-(z))+
\sum_{j=1}^{r}
(J_{x_j}^+(z)\Phi_{x_ju}(z)-\Phi_{x_ju}(z)
 J_{x_j}^-(z)),\tag 2.3\endgather
$$
\endproclaim

\demo{Proof} The logic of deduction is the same
as in the untwisted case (see, e.g.,\cite{FR}).
Equation (2.3) is nothing else but the intertwining
relation between $\Phi(z)$ and $\partial$:
$$
z\frac{d}{dz}\Phi_u(z)=-[\partial, \Phi_u(z)].\tag 2.4
$$
Substituting the expression for $\partial$ (Eq. (1.7)) into this relation,
after some calculations we obtain (2.4).
\enddemo

Now let us define the twisted correlation functions.
Let $V_1,...,V_N$ be highest weight representations of $\frak g$,
 and let
$\Phi^i_{u_i}(z_i): M_{\lambda_i,k}\to \hat M_{\lambda_{i-1},k}$, $1\le i\le n$
be intertwining operators. Set $\lambda_n=\lambda$ and $\lambda_0=\nu$.
Let $v_{\lambda}$ is the highest weight vector of $M_{\lambda,k}$, and
let $v_{\nu}^*$ be the lowest weight vector of the restricted dual module
to $M_{\nu,k}$. Consider the correlation function
$$
\Psi_{u_1,...,u_n}(z_1,...,z_n)=<v_{\nu}^*,\Phi^1_{u_1}(z_1)\dots\Phi^n_{u_n}(z_n)
v_{\lambda}>,\ u_i\in V_i, \tag 2.5
$$
which makes sense in the region $|z_1|>\dots>|z_n|$. The function $\Psi$ can
be regarded as taking values in the space $V_1\otimes\dots\otimes V_n$.

The function $\Psi$ satisfies a twisted version of the trigonometric
KZ equations.
Before writing these equations down, let us recall the definition of
the trigonometric classical $r$-matrix \cite{BeDr}.

Let
$\Omega=\sum_{\alpha\in\Delta^+}(e_{\alpha}\otimes f_{\alpha}+f_{\alpha}\otimes
e_{\alpha})+\sum_{j=1}^rx_j\otimes x_j$, $\Omega\in{\frak g}\otimes {\frak g}$.
For $0\le p\le N-1$, let ${\frak g}_p$ be the eigenspace of $\gamma$ in $\frak
g$ with the eigenvalue $e^{2\pi \text{i}p/N}$. Let $\Omega^p_{\gamma}$
be the orthogonal
projection of $\Omega$ to the subspace ${\frak g}_p\otimes {\frak g}_{-p}$ of
${\frak g}\otimes{\frak g}$.  The trigonometric
$r$-matrix has the form
$$
r(z)=\frac{\Omega^0_{\gamma}}{2}+\sum_{p=0}^{N-1}\frac{\Omega^p_{\gamma}z^p}
{z^N-1}.\tag 2.6
$$

Introduce the convenient notation: $$\gather
a_i=\text{Id}_{V_1\otimes\dots\otimes V_{i-
\text{Id}}}\otimes a\otimes 1_{V_{i+1}\otimes\dots\otimes V_n},\
a\in U({\frak g}),\
1\le i\le n;\\ \text {if }r=\sum a^s\otimes b^s,\ a^s,b^s\in U({\frak g}),\
\text{then } r_{ij}=\sum (a^s)_i(b^s)_j.\tag 2.7\endgather
$$

The main property of the trigonometric $r$-matrix $r(z)$ is the classical
Yang-Baxter equation:
$$
[r_{ij}(z_i/z_j),r_{jk}(z_j/z_k)]+
[r_{jk}(z_j/z_k),r_{ki}(z_k/z_i)]+
[r_{ki}(z_k/z_i),r_{ij}(z_i/z_j)]=0.\tag 2.8
$$

\proclaim{Theorem 2.4} The function $\Psi$ satisfies the following system of
differential equations:
$$
(k+1)z_i\frac{\partial \Psi}{\partial z_i}=\sum_{j\ne i}r_{ij}(z_i/z_j)\Psi+
\frac{1}{2}(h_{\lambda}+h_{\nu})\Psi.\tag 2.9
$$
\endproclaim

\demo{Proof} The proof is analogous to that in the untwisted case
(see \cite{FR}).
It is based on the direct use of relation (2.3).
\enddemo
\vskip .1in

Since the twisted intertwining operators are obtained from the usual
ones by a simple transformation, we should expect that system (2.9)
should reduce to the untwisted trigonometric KZ equations. This turns
out to be the case. Indeed, set $\zeta_j=z_j^N$, and
$\hat\Psi(\zeta_1,...,\zeta_n)=(z_1^{lh_{\rho}})_1\dots
(z_n^{lh_{\rho}})_n\Psi(z_1,...,z_n)$.
Then the function $\hat \Psi$
satisfies the trigonometric KZ equations:
$$
N(k+1)\zeta_i\frac{\partial \hat\Psi}{\partial \zeta_i}=\sum_{j\ne
i}\frac
{\Omega_{ij}^+\zeta_i+\Omega_{ij}^-\zeta_j}{\zeta_i-\zeta_j}\Psi+
\frac{1}{2}(h_{\hat\lambda}+h_{\hat\nu}+2h_{\rho})\hat\Psi,\tag 2.10
$$
where $\Omega^{\pm}$ are the half Casimir operators defined in the
introduction, and $\hat\lambda=\lambda+k\rho$, $\hat\nu=\nu+k\rho$.
Therefore, the standard theory of the KZ equations can be applied to
the study of the properties of (2.9).

\heading
{\bf 3. Traces of intertwiners and elliptic $r$-matrices}
\endheading

 From now on the letter $\frak g$ will denote the Lie algebra
${\frak sl}_N(\Bbb C)$ of traceless $N\times N$ matrices with complex
entries. The dual Coxeter number of this algebra is $N$, and the rank is $N-1$.
The Cartan subalgebra $\frak h$ is the subalgebra of diagonal matrices, and
the element $C$ is the matrix
$\text{diag}(1,\varepsilon^{-1},\varepsilon^{-2},...,\varepsilon^{-N+1})$
(up to a
factor).

Let $B$ be the $N\times N$ matrix of zeros and ones
corresponding to the cyclic permutation $(12...N)$. Note that $BC=\varepsilon
CB$.

Define a new inner automorphism $\beta$ of $\frak g$:
$\beta(a)=BaB^{-1}$, $a\in\frak g$.
This automorphism has order $N$ and
commutes with $\gamma$: $\beta\circ\gamma=\gamma\circ\beta$.

The action of the automorphism $B$ naturally extends to $\hat{\frak g}$ and
$\hat{\frak g}_{\gamma}$. Note that on $\hat{\frak g}_{\gamma}$, unlike
$\frak g$ and $\hat{\frak g}$, $B$ is an outer automorphism: it corresponds to
the rotation of the affine Dynkin diagram of $\hat{\frak g}$
(which is a regular $N$-gon)
through the angle $2\pi/N$.

The action of $\beta$ in $\hat{\frak g}_{\gamma}$ preserves degree, hence,
it preserves the polarization. Therefore, it transforms Verma modules into
Verma modules. In other words, we can regard $B$ as an operaator
$B:M_{\lambda,k}\to M_{\beta(\lambda),k}$, where by convention
$\beta(\lambda)(h)=\lambda(\beta^{-1}(h))$.  This operator intertwines the
usual action of $\hat{\frak g}_{\gamma}$
and the action twisted by $\beta$:
$\beta(a)Bw=Baw$, $a\in \hat{\frak g}_{\gamma}$, $w\in M_{\lambda,k}$.

Let $\nu=\beta^{-1}(\lambda)$, and let $\Phi^j_{u_j}(z_j)$ be as above (cf.
section 2). Let $q$ be a complex number, $0<|q|<1$. Following the idea of
Frenkel, Reshetikhin (\cite{FR}, Remark 2.3) and Bernard \cite{Ber},
introduce the following function:
$$
F_{u_1,...,u_n}(z_1,...,z_n|q)=\text{Tr}\mid
_{M_{\nu,k}}(\Phi^1_{u_1}(z_1)\dots\Phi^n_{u_n}(z_n)Bq^{-\partial}).\tag
3.1
$$
This function takes values in $V_1\hat\otimes\dots\hat\otimes V_N$. From now on
it will be the main object of our study.

It turns out that the $n$-point trace $F(z_1,...,z_n|q)$ defined by (3.1)
satisfies a remarkable system of differential equations involving
elliptic solutions of the classical Yang-Baxter equation for ${\frak sl}_N$.
Let us deduce these equations. The idea of the method of deduction is due to
Frenkel and Reshetikhin.

Differentiating by $z_j$, we get
$$
\gather
(k+1)\frac{\partial}{\partial z_j}F(z_1,...,z_n|q)=\\
\text{Tr}\left(\Phi^1_{u_1}(z_1)\dots\sum_{\alpha}(J_{e_{\alpha}}^+
(z_j)\Phi^j_{f_{ \alpha}u_j}(z_j)-\Phi^j_{f_{ \alpha}u_j}(z_j)J_{e_{\alpha}}^-
(z_j))\dots\Phi^n_{u_n}(z_n)Bq^{-\partial}\right)+\\
\text{Tr}\left(\Phi^1_{u_1}(z_1)\dots\sum_{\alpha}(J_{f_{\alpha}}^+
(z_j)\Phi^j_{e_{ \alpha}u_j}(z_j)-\Phi^j_{e_{ \alpha}u_j}(z_j)J_{f_{\alpha}}^-
(z_j))\dots\Phi^n_{u_n}(z_n)Bq^{-\partial}\right)+\\
\text{Tr}\left(\Phi^1_{u_1}(z_1)\dots\sum_{l=1}^{N-1}(J_{x_l}^+ (z_j)
\Phi^j_{x_l
u_j}(z_j)-\Phi^j_{x_lu_j}(z_j)J_{x_l}^-(z_j))\dots\Phi^n_{u_n}(z_n)Bq^{-
\partial}\right).\tag 3.2
\endgather
$$

Now let us pull the currents $J^+$, $J^-$ around: the currents $J^-$ will
move to the right up to the end, then jump at the beginning and continue to
move to the right, and so on; the currents $J^+$ will move to the left
up to the beginning, then jump at the end and continue to move to the left,
and so on. As we do these permutations, we will need relations (2.2)
and also the following identities:
$$
q^{-\partial}J_a^{\pm}(z)=qJ_a^{\pm}(qz)q^{-\partial},\tag 3.3
$$
$$
BJ_a^{\pm}(z)=J_{\beta(a)}^{\pm}(z)B.\tag 3.4
$$

After $J^+$ and $J^-$ have made $M$ complete circles, equation
(3.2) will have the form
$$
\gather
(k+1)\frac{\partial}{\partial z_j}F(z_1,...,z_n|q)=\\
\text{Tr}\biggl(\Phi^1_{u_1}(z_1)\dots\sum_{\alpha}(q^MJ_{\beta^M(e_
{\alpha})}^+
(q^Mz_j)\Phi^j_{f_{ \alpha}u_j}(z_j)-\\
\Phi^j_{f_{ \alpha}u_j}(z_j)q^{-M}
J_{\beta^{-M}(e_{\alpha})}^-(q^{-M}z_j))\dots\Phi^n_{u_n}(z_n)Bq^{-
\partial}\biggr)+\\
\text{Tr}\biggl(\Phi^1_{u_1}(z_1)\dots\sum_{\alpha}(q^MJ_{\beta^M(f_{\alpha})}
^+
(q^Mz_j)\Phi^j_{e_{ \alpha}u_j}(z_j)-
\\  \Phi^j_{e_{ \alpha}u_j}(z_j)q^{-M}
J_{\beta^{-M}(f_{\alpha})}^-(q^{-M}z_j))\dots\Phi^n_{u_n}(z_n)Bq^{-
\partial}\biggr)+\\
\text{Tr}\biggl(\Phi^1_{u_1}(z_1)\dots\sum_{l=1}^{N-1}(q^MJ_{\beta^M(x_l)}^+
(z_j)\Phi^j_{x_l u_j}(z_j)-\\
\Phi^j_{x_lu_j}(z_j)q^{-M}J_{\beta^{-M}(x_l)}^-
(q^{-M}z_j))\dots\Phi^n_{u_n}(z_n)Bq^{-\partial}\biggr)+\\
\sum_{i=1}^nX_{ij}^M
F(z_1,...,z_n|q),\tag 3.5
\endgather
$$
where
$$
X_{ij}^M=\cases \sum_{p=-M}^MX_{ij}^{M,p}& i\ne j\\
\sum_{p=-M,p\ne 0}^MX_{jj}^{M,p}& i= j\endcases
\tag 3.6
$$
and
$$
\gather
X_{ij}^{M,p}=\sum_{\alpha}\frac{q^p(q^pz_i)^{N-1-
|\alpha|}z_j^{|\alpha|}}{q^{pN}z_i^N-z_j^N}(\beta^p(e_\alpha))_i(f_{\alpha})_j
\\
+ \sum_{\alpha}\frac{q^p(q^pz_i)^{|\alpha|-1}z_j^{N-|\alpha|}}{q^{pN}z_i^N-
z_j^N}(\beta^p(f_\alpha))_i(e_{\alpha})_j
+ \sum_{l}\frac{1}{2z_i}\frac{(q^pz_i)^N+z_j^N}{q^{pN}z_i^N-
z_j^N}(\beta^p(x_l))_i(x_l)_j.  \tag 3.7
\endgather
$$

Now we want to pass to the limit $M\to\infty$. Right now we cannot do so since
the limit does not exist. In order to be able to pass to the limit,
we should write down equation (3.5) for $M=L,L+1,...,L+N-1$, add these $N$
equations together, and divide by $N$:
$$
\gather
(k+1)\frac{\partial}{\partial z_j}F(z_1,...,z_n|q)=\\
\frac{1}{N}\sum_{M=L}^{L+N-1}\biggl[
\text{Tr}\biggl(\Phi^1_{u_1}(z_1)\dots\sum_{\alpha}(q^MJ_{\beta^M(e_{\alpha})}
^+
(q^Mz_j)\Phi^j_{f_{ \alpha}u_j}(z_j)-
\\
\Phi^j_{f_{ \alpha}u_j}(z_j)q^{-M}
J_{\beta^{-M}(e_{\alpha})}^-(q^{-M}z_j))\dots\Phi^n_{u_n}(z_n)Bq^{-
\partial}\biggr)+\\
\text{Tr}\biggl(\Phi^1_{u_1}(z_1)\dots\sum_{\alpha}(q^MJ_{\beta^M(f_{\alpha})}
^+
(q^Mz_j)\Phi^j_{e_{ \alpha}u_j}(z_j)-
\\
\Phi^j_{e_{ \alpha}u_j}(z_j)q^{-M}
J_{\beta^{-M}(f_{\alpha})}^-(q^{-M}z_j))\dots\Phi^n_{u_n}(z_n)Bq^{-
\partial}\biggr)+\\
\text{Tr}\biggl(\Phi^1_{u_1}(z_1)\dots\sum_{l=1}^{N-1}(q^MJ_{\beta^M(x_l)}^+
(z_j)\Phi^j_{x_l u_j}(z_j)-\\
\Phi^j_{x_lu_j}(z_j)q^{-M}J_{\beta^{-M}(x_l)}^-
(q^{-M}z_j))\dots\Phi^n_{u_n}(z_n)Bq^{-\partial}\biggr)+\\
\sum_{i=1}^nX_{ij}^M\biggr] F(z_1,...,z_n|q),\tag 3.8
\endgather
$$

In the obtained equation it will already be possible to take the limit.
Moreover, since $\sum_{p=0}^{N-1}\beta^p(h)=0$ for $h\in {\frak h}$, the part
of the right hand side of (3.8) involving currents associated with the
Cartan
subalgebra elements will disappear as
$M\to\infty$. The same will happen to the currents associated with the
root elements $e_{\alpha}$, $f_{\alpha}$ because these currents do not
contain a term of degree $-1$ which is the only term that could
possibly have given a
nonzero limit. Thus, in the limit we get a simple equation:
$$
\frac{\partial}{\partial z_j}F(z_1,...,z_n|q)=
\left(\sum_{i=1}^nX_{ij}^{\infty}\right)F(z_1,...,z_n|q),\tag 3.9
$$
where $X_{ij}^{\infty}=\lim_{L\to\infty}\frac{1}{N}\sum_{M=L}^{L+N-
1}X_{ij}^M$.

The function $X_{ij}^{\infty}$ admits a simple description in terms of
elliptic functions.

First of all, it is easy to check that $X_{jj}^{M}=0$ for any $M$.

Next, looking at the function $X_{ij}^M$ can be represented in the form
$X_{ij}^M=\frac{1}{z_i}\rho^M_{ij}(z_i/z_j)$, where
$\rho^M(z)$ is a rational function with values in ${\frak g}\otimes {\frak
g}$:
$$
\gather
\rho^M(z)=\sum_{p=-M}^M\biggl(\sum_{\alpha\in\Delta^+}\frac{(q^pz)^{N-
|\alpha|}}{q^{pN}z^N-1}f_{\alpha}\otimes
\beta^p(e_{\alpha})+\\
\sum_{\alpha\in\Delta^+}
\frac{(q^pz)^{
|\alpha|}}{q^{pN}z^N-1}e_{\alpha}\otimes
\beta^p(f_{\alpha})+\sum_{l=1}^{N-1}\frac{q^{pN}z^N+1}{2(q^{pN}z^N-
1)}x_l\otimes \beta^p(x_l)\biggr)\tag 3.10\endgather
$$

Therefore, $X_{ij}^{\infty}=\frac{1}{z_i}\rho_{ij}(z_i/z_j|q)$, where
$\rho(z|q)$
is an elliptic function of $\log z$ with values in ${\frak g}\otimes {\frak
g}$ (we have used notation (2.7)).  We can tell what this function
looks like
 by looking at its residues.

 From (3.10) we see that the only poles of $\rho(z|q)$ are at the points
$\varepsilon^mq^p$, and all these poles are simple. The residue of
$\rho(z|q)$
 at
$z=\varepsilon^mq^p$ is equal to that of $\rho^M(z)$ for $M\ge |p|$, i.e. it
equals $$
\gather
\text{Res}\mid_{z=\varepsilon^mq^p}\rho(z|q)=\frac{q^p
}{N}\biggl(\sum_{\alpha\in\Delta ^+}(\varepsilon^{-
m|\alpha|}f_{\alpha}\otimes\beta^p(e_{\alpha})+
\\
\varepsilon^{m|\alpha|}e_{\alpha}
\otimes \beta^p(f_{\alpha})+\sum_{l=1}^{N-1}x_l\otimes\beta^p(x_l)\biggr)=
\frac{q^p}{N}(1\otimes \gamma^{-m}\beta^p)(\Omega)\tag 3.11
\endgather
$$

Because of the obvious homogeneity property $\sum_{i}z_i\frac{\partial
F}{\partial z_i}=0$, we have the unitarity relation
$\rho_{ij}(z)=-\rho_{ji}(-z)$.

Let $q=e^{2\pi \text{i}\tau}$. Let
$$
\zeta(x|\tau)=\frac{1}{x}+\lim_{M\to \infty}\sum_{-M\le j,l\le
M,j^2+l^2>0}
\biggl[\frac{1}{x-j-l\tau}+\frac{x}{(j+l\tau)^2}\biggr]\tag 3.12
$$
be the standard Weierstrass function.
Equation (3.11) implies that
$$
\gather
\rho(z|q)=\frac{\Omega}{2\pi\text{i}N}\zeta\biggl(\frac{\log z}{2\pi
\text{i}}|N\tau\biggr)+ \\
\frac{1}{2\pi\text{i}N}\sum_{0\le m,p\le N-1,m^2+n^2>0}(1\otimes \gamma^{-
m}\beta^p)(\Omega)\biggl[\zeta\biggl(\frac{\log z}{2\pi
\text{i}}-\frac{m}{N}
-p\tau|N\tau\biggr)-\zeta\biggl(-\frac{m}{N}-p\tau|\tau\biggr)\biggr].\tag
3.13\endgather
$$

Thus, we have proved
\proclaim {Theorem 3.1} The function $F(z_1,...,z_n|q)$ satisfies the system
of differential equations
$$
(k+1)z_i\frac{\partial F}{\partial z_i}=\sum_{j\ne
i}\rho_{ij}(z_i/z_j|q)F,\ 1\le i\le n.\tag
3.14
$$
\endproclaim

The function $\rho(z|q)$ is a solution of the classical Yang-Baxter equation.
It is easy to see that this function is nothing else but the elliptic
$r$-matrix for ${\frak sl}_N$ due to A.Belavin \cite{Be}. Thus, we have given
a representation-theoretical interpretation of the local system associated
with the elliptic solutions of the classical Yang-Baxter equation.

{\bf Remark. } A.Belavin and V.Drinfeld \cite{BeDr} showed that elliptic
solution exist only for the simple Lie algebra ${\frak sl}_N$, and
every nondegenerate elliptic solution is equivalent (=conjugate) to
$\text{const}\cdot \rho(z|q)$ (for a suitable primitive $N$-th root of
unity $\varepsilon$).

Observe that equations (3.14) transform into the KZ
equations (2.9) as $q\to 0$.
This was to be expected since $\lim_{q\to
0}q^{-\frac{<\lambda,\lambda>}
{2(k+1)}}F(\bold z|q)=\Psi(\bold z)$.
We will call equations (3.14) {\it the elliptic Knizhnik-Zamolodchikov (KZ)
equations}.

There is one more equation satisfied by the function $F$ - a differential
equation involving the first derivative by $q$. Differentiating $F$ by $q$, we
obtain
$$
-q\frac{\partial F}{\partial q}= \text{Tr}\mid
_{M_{\lambda,k}}(\Phi^1_{u_1}(z_1)\dots\Phi_{u_n}(z_N)Bq^{-\partial}\partial
).\tag 3.15
$$
Plugging the expression for $\partial$ in (3.15), we see that in order to
obtain the differential equation for $F$, it would suffice to express
the traces
$$
\gather
\text{Tr}\mid
_{M_{\lambda,k}}(\Phi^1_{u_1}(z_1)\dots\Phi^n_{u_n}(z_N)Bq^{-
\partial}f_{\alpha}\otimes t^{-|\alpha|-mN}e_{\alpha}\otimes t^{|\alpha|+mN}
),\\
\text{Tr}\mid
_{M_{\lambda,k}}(\Phi^1_{u_1}(z_1)\dots\Phi^n_{u_n}(z_N)Bq^{-
\partial}e_{\alpha}\otimes t^{|\alpha|-mN}f_{\alpha}\otimes t^{-|\alpha|+mN} )
,\\
\text{Tr}\mid
_{M_{\lambda,k}}(\Phi^1_{u_1}(z_1)\dots\Phi^n_{u_n}(z_N)Bq^{-
\partial}x_l\otimes t^{-mN}x_l\otimes t^{mN} ) \tag 3.16
\endgather
$$
in terms of the original trace $F(\bold z|q)$. This can be done as follows.

Take the expression for the first trace in (3.16) and move the factor
$e_{\alpha}\otimes t^{|\alpha|+mN}$ from left to right. When it has
made $N$ full circles, we will have
$$
\gather
\text{Tr}\mid
_{M_{\lambda,k}}(\Phi^1_{u_1}(z_1)\dots\Phi^n_{u_n}(z_N)Bq^{-
\partial}f_{\alpha}\otimes t^{-|\alpha|-mN}e_{\alpha}\otimes t^{|\alpha|+mN}
)=\\
\sum_{j=1}^n\sum_{p=0}^{N-1}
z_j^{|\alpha|+mN}q^{-p(|\alpha|+mN)}\text{Tr}\mid
_{M_{\lambda,k}}(\Phi^1_{u_1}(z_1)\dots\Phi^j_{\beta^{-p}(e_{\alpha})u_j}(z_j)
\dots
\Phi^n_{u_n}(z_N)Bq^{-
\partial}f_{\alpha}\otimes t^{-|\alpha|-mN}
)+\\
q^{-N(|\alpha|+mN)}\text{Tr}\mid
_{M_{\lambda,k}}(\Phi^1_{u_1}(z_1)\dots\Phi^n_{u_n}(z_N)Bq^{-
\partial}(h_{\alpha}+(|\alpha|+mN)k)
)+\\
q^{-N(|\alpha|+mN)}\text{Tr}\mid
_{M_{\lambda,k}}(\Phi^1_{u_1}(z_1)\dots\Phi^n_{u_n}(z_N)Bq^{-
\partial}f_{\alpha}\otimes t^{-|\alpha|-mN}e_{\alpha}\otimes t^{|\alpha|+mN}
).\tag 3.17\endgather
$$
It is easy to see that for any $h\in \frak h$
$$
\text{Tr}\mid
_{M_{\lambda,k}}(\Phi^1_{u_1}(z_1)\dots\Phi^n_{u_n}(z_N)Bq^{-
\partial}h
)=-(1-\beta^{-1})^{-1}(h)F(\bold z|q).\tag 3.18
$$
Similarly we have
$$
\text{Tr}\mid
_{M_{\lambda,k}}(\Phi^1_{u_1}(z_1)\dots\Phi^n_{u_n}(z_N)Bq^{-
\partial}f_{\alpha}\otimes t^{-|\alpha|-mN}
)=-\sum_{j=1}^nz_j^{-|\alpha|-mN}(1-\beta^{-1} q^{|\alpha|+mN})^{-1}
(f_{\alpha})_jF(\bold z|q).\tag
3.19
$$
Thus, we obtain the following expression for the first trace in
(3.16):
$$
\gather
\text{Tr}\mid
_{M_{\lambda,k}}(\Phi^1_{u_1}(z_1)\dots\Phi^n_{u_n}(z_N)Bq^{-
\partial}f_{\alpha}\otimes t^{-|\alpha|-mN}e_{\alpha}\otimes t^{|\alpha|+mN}
)=\\
\sum_{i,j=1}^nz_i^{-|\alpha|-mN}z_j^{|\alpha|+mN}(1-\beta^{-1}q^{|\alpha|+mN})
^{-1}(f_{\alpha})_i(1-\beta^{-1}q^{-|\alpha|-mN})^{-1}(e_{\alpha})_jF(\bold
z|
q)+\\
(1-q^{-N(|\alpha|+mN)})^{-1}\biggl((1-\beta^{-1})^{-1}(-h_{\alpha})+
(|\alpha|+mN)k\biggr)F(\bold z|q).\tag 3.20\endgather
$$

The second and the third trace in (3.16) are treated quite similarly,
and finally after some calculations we obtain:
$$
(k+1)q\frac{\partial F}{\partial
q}=\sum_{i,j=1}^nL_{ij}\biggl(\frac{z_i}{z_j}\bigg|q\biggr)F,\tag 3.21
$$
where
$$
\gather
L(z|q)=\sum_{\alpha\in\Delta^+}\sum_{m\ge 0}z^{-|\alpha|-mN}
(1-\beta^{-1}q^{|\alpha|+mN})^{-1}(f_{\alpha})\otimes(1-\beta^{-1}
q^{-|\alpha|-mN})^{-1}(e_{\alpha})+\\
\sum_{\alpha\in\Delta^+}\sum_{m>0}z^{|\alpha|-mN}
(1-\beta^{-1}q^{-|\alpha|+mN})^{-1}(e_{\alpha})\otimes(1-\beta^{-1}
q^{|\alpha|-mN})^{-1}(f_{\alpha})+\\
\sum_{p=1}^{N-1}\sum_{m>0}z^{-mN}
(1-\beta^{-1}q^{mN})^{-1}(x_p)\otimes(1-\beta^{-1}q^{-mN})^{-1}(x_p)+\\
\frac{1}{2}\sum_{p=1}^{N-1}(1-\beta^{-1})^{-1}(x_p)\otimes
(1-\beta^{-1})^{-1}(x_p)
\tag
3.22\endgather
$$
is a function with values in ${\frak g}\otimes {\frak g}$
(once again, we use notation (2.7)). Note that
$L_{ij}=L_{ji}$. If $i=j$,
then by $L_{ii}$ we mean $\mu(L)_i$, where $\mu:U({\frak g})\otimes
U({\frak g})\to U({\frak g})$ is multiplication: $\mu(a\otimes b)=ab$.

 The form of equation (3.21) can be simplified. Indeed,
the consistency of (3.21) and (3.14) for all $k$ implies that
$$
\gather
q\frac{\partial
\rho(z|q)}{\partial q}=z\frac{\partial
}{\partial z}(L(z|q)+L(z^{-1}|q)),\text{ i.e.}\\
L(z|q)+L(z^{-1}|q)=2L(1,q)+\int_1^z\frac{q}{z}\frac{\partial
\rho(z|q)}{\partial q}dz.\tag 3.23\endgather
$$
Therefore, we finally get the theorem (in the formulation we use
notation 2.7)):

\proclaim{Theorem 3.2}
The function $F$ satisfies the differential equation
$$
\gather
(k+1)q\frac{\partial F}{\partial
q}=\sum_{i,j=1}^nL_{ij}(1|q)F+\sum_{i<j}s\biggl(\frac{z_i}{z_j}\bigg|
q\biggr)_{ij}F,\\
s(z|q)=\int_1^z\frac{q}{z}\frac{\partial
\rho(z|q)}{\partial q}dz
\tag 3.24\endgather
$$
\endproclaim

\proclaim{Corollary}
$$
\text{Tr}\mid_{M_{0,k}}(Bq^{-\partial})=1.\tag 3.25
$$
\endproclaim

\demo{Proof} Apply Theorem 3.2 to the case of a single intertwining
operator $\Phi(z):M_{0,k}\to M_{0,k}\otimes V^0_C(z)$, where $V^0$ is
the trivial representation of $\frak g$. In this case it is obvious
that $\Phi(z)=\text{Id}$. Therefore, (3.23) simply follows from the
fact that (3.22) acts by zero in $V^0$.
\enddemo
\vskip .1in

\heading
{\bf 4. Modular invariance of the elliptic KZ equations}
\endheading

Let us now discuss the modular invariance of the elliptic KZ
equations. Introduce new variables $y_j=\frac{N\log z_j}{2\pi \text{i}}$,
$\tau=\frac{\log q}{2\pi \text{i}}$. From now on let us use the
notation $\kappa=k+1$. Rewrite equations (3.14) and (3.24) in
the new coordinates:
$$
\gather
\kappa{\frac{\partial F}{\partial y_i}}=\sum_{j\ne
i}\rho^*_{ij}(y_i-y_j|\tau)F,\ 1\le i\le n\tag 4.1\\
\kappa{\frac{\partial F}{\partial
\tau}}=\sum_{i,j=1}^nL_{ij}^*(\tau)F+
\sum_{j<i}s^*_{ij}(y_i-y_j|\tau)F,\tag 4.2\endgather
$$
where
$$
\gather
\rho^*(y|\tau)=\frac{2\pi \text{i}}{N}\rho(z|q),\\
s^*(y|\tau)=2\pi \text{i}s(z|q)=\int_0^y\frac{\partial \rho^*(x|\tau)}{\partial
\tau}dx,\\
L^*(\tau)=2\pi \text{i}L(1|q).\tag 4.3
\endgather
$$

Let $\Gamma(N)$ be the congruence subgroup in $SL_2(\Bbb Z)$ consisting
of the matrices equal to the identity modulo $N$.
We have the following almost obvious property.

\proclaim{Proposition 4.1} Equations (4.1) are invariant with respect
to the group $\Gamma(N)$. That is, they are preserved under the change
of variables $\hat y_i=\frac{y_i}{c\tau+d}$, $\hat \tau=
\frac{a\tau+b}{c\tau+d}$ if $A=\biggl(\matrix a&b\\ c&d\endmatrix\biggr)\in
\Gamma(N)$.
\endproclaim

\demo{Proof}
Partial derivatives with respect to the new coordinates are given by
$$
\gather
\frac{\partial}{\partial \hat y_i}=(c\tau+d)\frac{\partial}{\partial y_i},\\
\frac{\partial}{\partial\hat\tau}=(c\tau+d)^2\frac{\partial}{\partial\tau}+
\sum_i c(c\tau+d)y_i\frac{\partial}{\partial y_i}.\tag 4.4
\endgather
$$
Thus we only have to prove that
$\rho^*(\hat y|\hat \tau)=(c\tau+d)\rho^*(y|\tau)$. To do this, it
is enough to observe that both sides of this equation

(i) have simple poles at
the points of the lattice $m+p\tau$ with the same residues\linebreak
$\frac{1}{N}(1\otimes\gamma^{-m}
\beta^p)(\Omega)$, and

(ii) satisfy the unitarity condition:
$\rho^*_{12}(y|\tau)=-\rho^*_{21}(-y|\tau)$.
\enddemo

Now let us study the invariance properties of equation (4.2).
First of all, it is easy to see that
$$
\frac{\partial\rho^*(\hat y|\hat\tau)}{\partial\hat\tau}=
(c\tau+d)^3\frac{\partial\rho^*(y|\tau)}{\partial\tau}+
c(c\tau+d)^2\frac{\partial \bigl(y\rho^*(y|\tau)\bigr)}{\partial y}.\tag 4.5
$$
Integrating this equation against $dy$, we get
$$s^*(\hat y,\hat
\tau)=(c\tau+d)^2s^*(y|\tau)+c(c\tau+d)\biggl(y\rho^*(y|\tau)-\lim_{x\to
0}x\rho^*(x|\tau)\biggr).\tag 4.6$$

By our definition, $\lim_{x\to
0}x\rho^*(x|\tau)=\frac{\Omega}{N}$.
Thus, under the change of variable $\tau\to\hat\tau,\ y\to\hat y$ equation
(4.2) transforms into the following equation:
$$
\gather
(c\tau+d)^2\kappa\frac{\partial F}{\partial
\tau}+c(c\tau+d)\sum_i\kappa y_i\frac{\partial F}{\partial y_i}=\\
\sum_{i<j}(c\tau+d)^2s^*(y_i-y_j|\tau)F+\sum_{i<j}c(c\tau+d)\biggl((y_i-y_j)
\rho^*(y_i-y_j|\tau)-\frac{\Omega_{ij}}{N}\biggr)F+\sum_{i,j}L_{ij}(\hat\tau)
F.\tag 4.7
\endgather
$$
Combining (4.7) with (4.1), we can get rid of the derivatives by $y_i$
and reduce (4.7) to the form:
$$
\kappa\frac{\partial F}{\partial
\tau}=\sum_{i<j}\biggl(s_{ij}^*(y|\tau)-\frac{c\Omega_{ij}}{(c\tau+d)N}\biggr)
F+
(c\tau+d)^{-2}\sum_{i,j}L^*_{ij}(\hat\tau)F.
\tag 4.8
$$
Now we need to find the law of transformation of $L^*(\tau)$.
\proclaim{Lemma}
$$
L^*(\hat\tau)=(c\tau+d)^2L^*(\tau)+\frac{c(c\tau+d)\Omega}{2N}.\tag 4.9
$$
\endproclaim

\demo{Proof} Let
$C(\tau)=L^*(\hat\tau)-(c\tau+d)^2L^*(\tau)-\frac{c(c\tau+d)
\Omega}{2N}\in {\frak
g}\otimes\frak g$. We know that both equations (4.2) and (4.8)
are consistent with (4.1). Therefore, we have
$$
[C_{11}(\tau)+2C_{12}(\tau)+C_{22}(\tau)+\frac{c(c\tau+d)(\Omega_{11}+
\Omega_{22})}{N},\rho^*_{12}(y
|\tau)]=0.\tag 4.10
$$
Observe that $\Omega_{11}=\Delta\otimes 1$, $\Omega_{22}=1\otimes\Delta$,
where $\Delta\in U(\frak g)$ is the Casimir element.
Since the Casimir element commutes with the Lie algebra action, (4.10)
reduces to the relation
$$
[C_{11}(\tau)+2C_{12}(\tau)+C_{22}(\tau),\rho^*_{12}(y
|\tau)]=0.\tag 4.11
$$
This relation has to hold for all $y$, which implies that
the expression $\tilde C(\tau)=
C_{11}(\tau)+2C_{12}(\tau)+C_{22}(\tau)$ commutes with all
$\gamma$-invariant elements in ${\frak g}\otimes{\frak g}$, i.e with all
elements of the form $x_i\otimes  x_j$,
$e_{\alpha}\otimes\beta^p(f_{\alpha})$
 and $f_{\alpha}\otimes\beta^p(e_{\alpha})$.

We are going to prove that $C(\tau)=0$.
Let
$$C(\tau)=\sum_i a_i(\tau)x_i\otimes x_i+\sum_{\alpha,p}b_{\alpha
p}(\tau)(e_{\alpha}\otimes \beta^p(f_{\alpha})+\beta^p(f_{\alpha})\otimes
e_{\alpha}).\tag 4.12
$$

Pick arbitrary two elements $X$ and $Y$ in $\frak h$. Consider the
expression
$$
\gather
C_{XY}=\bigl[X\otimes 1,[1\otimes Y,2C]\bigr]=\bigl[X\otimes
1,[1\otimes Y,\tilde C]\bigr]=\\
-2\sum_{\alpha,p}b_{\alpha
p}(\alpha(X)\alpha(\beta^{-p}(Y))e_{\alpha}\otimes\beta^p(f_{\alpha})+
\alpha(Y)\alpha(\beta^{-p}(X))f_{\alpha}\otimes
\beta^p(e_{\alpha})\tag 4.13\endgather
$$
(for brevity we do not specify explicitly the dependence on $\tau$).
This expression has to commute with $x_i\otimes x_j$ for all $i,j$.
This immediately implies that $C_{XY}=0$ for all $X,Y$, i.e.
$b_{\alpha p}=0$ for all $\alpha,p$. Therefore,
$$
C=\sum_i a_i\cdot x_i\otimes x_i.\tag 4.14
$$
Let $X_i=1\otimes x_i+x_i\otimes 1$. Then
$\tilde C=\sum_ia_iX_i^2$. Therefore,
$$
\gather
[\tilde
C,e_{\alpha}\otimes\beta^p(f_{\alpha})]=
\\
\sum_{i}\alpha((1-\beta^{-p})(x_i))
a_i(X_i
e_{\alpha}\otimes\beta^p(f_{\alpha})+e_{\alpha}\otimes\beta^p(f_{\alpha})X_i).
\tag 4.15\endgather
$$
In order for (4.15) to be zero for any $\alpha,p$, we must have
$\sum_{i}\alpha((1-\beta^{-p})(x_i))
a_iX_i=0$ for all $\alpha$, $p$, which can only happen when all $a_i$
are zero (because the roots span the space $\frak h^*$) Q.E.D.
\enddemo

The Lemma we just proved implies
\proclaim{Proposition 4.2} Under the change of variables $(y,\tau)\to (\hat
y,\hat\tau)$ system of equations (4.1),  (4.2) transforms into a system
of equations equivalent to the combination of
(4.1) and the following equation:
$$
\kappa{\frac{\partial F}{\partial
\tau}}=\sum_{i,j=1}^nL_{ij}^*(\tau)F+
\sum_{j<i}s^*_{ij}(y_i-y_j|\hat \tau)F+\sum_{i=1}^n\frac{c\Delta_i}
{2N(c\tau+d)}F,\tag 4.16
$$
where $\Delta_i$ is the Casimir operator in the $i-th$ factor of the tensor
product $V_1\hat\otimes V_2\hat\otimes\dots\hat\otimes V_n$.
\endproclaim

Thus, the system of equations (4.1), (4.2) is almost invariant under
$\Gamma(N)$: the first $n$ equations are unchanged under the action of this
group whereas the last equation gets a very simple extra term.

Consider {\it the fundamental solution}
$\Cal F(\bold y|\tau)$ of the system (4.1), (4.2) -- a solution with
values in $\text{End}(V_1\hat\otimes\dots\hat\otimes V_n)$
determined by the initial condition
$$
\Cal F(y_1,...,y_n|\tau)\to \text{Id}\text{ as } \tau \to
\text{i}\infty, \text{Im}y_{j+1}-\text{Im}y_{j}\to +\infty\text{ for
} 1\le j\le n-1.\tag 4.18
$$

\proclaim{Theorem 4.3}(on the modular invariance of solutions of the
elliptic KZ equations).
$$
\Cal F(\hat y_1,...,\hat y_n|\hat\tau)= (c\tau+d)^{\frac{1}{2N\kappa}
\sum_i\Delta_i}\Cal F(y_1,...,y_n|\tau)\chi(A),\tag 4.19
$$
where
$\chi(A)$ is an operator in $V_1\hat\otimes
V_2\hat\otimes\dots\hat\otimes V_n$ dependent only on the $2\times 2$
matrix $A$ (i.e. independent of $y_i$ and $\tau$).
\endproclaim

Assume that the representations $V_j$ are irreducible. Then $\Delta_j$ are
simply complex numbers. In this case the function $\chi(A)$ is a projective
representation of $\Gamma(N)$:
$$
\chi(A_1)\chi(A_2)=\sigma(A_1,A_2)\chi(A_1A_2).\tag 4.20
$$

The 2-cocycle $\sigma(A_1,A_2)$ is very easy to describe. Let
$A=\biggl( \matrix a&b\\ c&d\endmatrix\biggr)\in
SL_2(\Bbb Z)$ be called positive if either $c>0$ or if $c=0$ but
$d>0$; otherwise, let $A$ be called negative. Clearly, $A$ is positive
if and only if $-A$ is negative. Let $|A|$ denote $A$ if $A$ is
positive and $-A$ if $A$ is negative. Then
$$
\sigma(A_1,A_2)=\cases 1,&|A_1||A_2| \text{is positive}\\
e^{\frac{\pi\text{i}\sum_j
\Delta_j}{2N\kappa}},& |A_1||A_2| \text{is negative}\endcases\tag 4.21
$$
Of course, this cocycle is a coboundary since $H^2(\Gamma(N),\Bbb
Z)=0$, so the action of $\Gamma(N)$ in
the projectivization of $V_1\hat\otimes
V_2\hat\otimes\dots\hat\otimes V_n$ comes from a linear action.

 Thus, the theory of the elliptic KZ
equations gives us a natural method to assign to every set of irreducible
finite dimensional
representations $V_1,...,V_n$ of ${\frak sl}_N$ an action of the
congruence subgroup $\Gamma(N)$ of $SL_2(\Bbb Z)$ in the tensor product
of these representations, $V_1\otimes\dots\otimes V_n$.

In fact this construction allows us to obtain
 a representation of the entire modular
group $SL_2(\Bbb Z)$. For this we need to assume that $V_i$ are finite
dimensional representations of $GL_N(\Bbb C)$ for all $i$.
For brevity we will also assume that $N$ is odd (this
assumption is not very essential, but in the even case one has to be a
little bit more careful).

We will need to use the Weil representation of the group $SL_2(\Bbb
Z/N\Bbb Z)$. This representation is defined as follows. Take the
$N$-dimensional vector space $U=\Bbb C^N$ and define an action of the
Heisenberg group $H_N=<x,y|x^N=y^N=1,xyx^{-1}y^{-1}\text{ commutes
with $x,y$ }>$ in this space by $x\to B$, $y\to C$. This is the basic
irreducible representation of $H_N$.
Now observe that $SL_2(\Bbb Z/N\Bbb Z)$ acts
by automorphisms of $H_N$: if $A=\biggl(\matrix a&b\\
c&d\endmatrix\biggr)$ then $A(x)=x^ay^b,\ A(y)=x^cy^d$.
Moreover, the representation $U^A$ of $H_N$ obtained from $U$ by
twisting of $U$ by $A$ is isomorphic to $U$. Therefore, the group
$SL_2(\Bbb Z/N\Bbb Z)$ projectively acts in $U$ in such a way that
$Azu=A(z)Au$, $A\in SL_2(\Bbb Z/N\Bbb Z)$, $z\in H_N$, $u\in PU$, and
this action is unique. The space $U$ with the constructed projective action of
$SL_2(\Bbb Z/N\Bbb Z)$ is referred to as the Weil representation.
This representation defines a homomorphism $W_0:SL_2(\Bbb Z/N\Bbb Z)\to
PGL_N(\Bbb C)$. Since the group $SL_2(\Bbb Z/N\Bbb Z)$ for odd $N$
does not have nontrivial central extensions, this homomorphism lifts
to a map $W: SL_2(\Bbb Z/N\Bbb Z)\to GL_N(\Bbb C)$.

\proclaim{Proposition 4.4} Let $A=\biggl(\matrix a&b\\ c&d \endmatrix
\biggr)\in SL_2(\Bbb Z)$, and let the change of variables $(y,\tau)\to
(\hat y,\hat\tau)$ be defined as in Proposition 4.1. Then
$$
\Cal F(\hat y_1,...,\hat y_n|\hat \tau)=
(c\tau+d)^{\frac{1}{2N\kappa}
\sum_i\Delta_i}\theta(A)\Cal F(y_1,...,y_n|\tau)\chi(A),\tag 4.22
$$
where $\chi(A)$ is a projective representation of $SL_2(\Bbb Z)$ in
$V_1\otimes\dots\otimes V_N$, and $\theta$ is the composition of
the three maps:
$$SL_2(\Bbb Z) @>\text{mod }N>> SL_2(\Bbb Z/N\Bbb Z)@>W>> GL_N(\Bbb
C)@>\pi_1\otimes \dots\otimes \pi_n>>
\text{Aut}(V_1\otimes\dots\otimes V_N),\tag 4.23$$
where $\pi_j: GL_N\to \text{Aut}(V_j)$ are the homomorphisms defining the
action of the group $GL_N$ in $V_i$.
\endproclaim

The proof of this statement is simple and similar to the proof of
Theorem 4.3. The Weil representation of $SL_2(\Bbb Z/N\Bbb Z)$ arizes
naturally when we consider the transformations of the lattice $L$ of
poles of $\rho^*(y|\tau)$ which do not preserve the lattice of periods
$NL$.

The function $\chi(A)$ satisfies equation (4.20) with the 2-cocycle
$\sigma$ still being
defined by (4.21) (now for the entire $SL_2(\Bbb Z)$).
Since $H^2(SL_2(\Bbb Z),\Bbb Z)=0$, this cocycle is
a coboundary, so $\chi(A)$ comes from a linear action of $SL_2(\Bbb Z)$.

{\bf Remark. }
It is not clear how to compute the representation
$\chi(A)$ for any nontrivial example. A good example to start with
would be $N=2$, $n=1$, and $V_1$ is the 4-dimensional irreducible
representation of ${\frak sl}_2$. In this case it seems that $V$ will
be a direct sum of two irreducible 2-dimensional
representations of $\Gamma(2)$, and the solutions of (3.24) will be
some nontrivial vector-valued modular functions.
\vskip .1in

\heading
{\bf 5. Some examples}
\endheading

Let us consider the special case $N=2$, ${\frak g=sl}_2$. In this case
$\beta$ acts as follows: $\beta(e)=f$, $\beta(f)=e$, $\beta(h)=-h$.
Therefore, we have
$$
\rho(z|q)=\frac{1}{2}\biggl[a(z|q)(e\otimes e+f\otimes f)+b(z|q)(e\otimes
f+f\otimes e)+c(z|q)h\otimes h\biggr],\tag 5.1
$$
where
$$
\gather
a(z|q)=\frac{1}{2\pi\text{i}}\biggl[\zeta\biggl(\frac{\log z}{2\pi
\text{i}}-\frac{1}{2}|2\tau\biggr)-\zeta\biggl(\frac{\log z}{2\pi
\text{i}}-
\frac{1}
{2}-\tau|2\tau\biggr)\biggr]+a_0(q),\\
b(z|q)=\frac{1}{2\pi\text{i}}\biggl[\zeta\biggl(\frac{\log z}{2\pi
\text{i}}|2\tau\biggr)-\zeta\biggl(\frac{\log
z}{2\pi \text{i}}-\tau|2\tau\biggr)\biggr]+b_0(q),\\
c(z|q)=\frac{1}{4\pi\text{i}}\biggl[\zeta\biggl(\frac{\log z}{2\pi
\text{i}}|2\tau\biggr)+\zeta\biggl(\frac{\log
z}{2\pi \text{i}}-\tau|2\tau\biggr)-\zeta\biggl(\frac{\log z}{2\pi
i}-\frac{1}{2}|2\tau\biggr)-\zeta\biggl(\frac{\log z}{2\pi \text{i}}-
\frac{1}{2}-\tau|2\tau\biggr)\biggr]+c_0(q),\tag
5.2\endgather
$$
and the constants $a_0,\ b_0,\ c_0$ are chosen to satisfy the
condition $\rho(z|q)=-\rho(-z|q)$.
We also get from (3.22)
$$
\gather
L(1|q)=-\sum_{m\ge
0}\frac{q^{2m+1}(1+q^{4m+2})}{(1-q^{4m+2})^2}
(e\otimes e+f\otimes f)\\
-\sum_{m\ge
0}\frac{2q^{4m+2}}{(1-q^{4m+2})^2}(e\otimes
f+f\otimes e)+\\
\frac{1}{2}\biggl(\frac{1}{8}+\sum_{m>
0}\frac{q^{2m}}{(1+q^{2m})^2}\biggr)
h\otimes h.\tag
5.3\endgather
$$

Now let us calculate some nontrivial 1-point traces.

{\bf Example 1. } Let $\Phi(z):M_{-\lambda,k}\to
\hat M_{\lambda,k}\otimes V^1_C(z)$ be an intertwining operator,
where $V^1$ is the two-dimensional irreducible representation of
${\frak sl}_2$. Then we must have $\lambda=\pm\frac{1}{2}$.
Denote the corresponding operators by $\Phi^{\pm}$, and introduce the notation:
$T_{\pm}(q)=\text{Tr}(\Phi^{\pm}Bq^{-\partial})$. It is clear that
these traces do not depend on $z$. Let us compute them.

We have proved that $T_{\pm}(q)$ satisfy the equation
$$
\kappa q\frac{\partial F}{\partial q}=L(1|q)F.\tag
5.4
$$
In the case we are considering the matrix $L$ turns out to be a scalar
$2\times 2$ matrix:
$$
L(1,q)\mid_{V^1}=\frac{1}{16}+
\frac{1}{2}\sum_{m> 0}\frac{q^{2m}}{(1+q^{2m})^2}
-\sum_{m\ge
0}\frac{2q^{4m+2}}{(1-q^{4m+2})^2}.\tag 5.5
$$
Therefore, we can explicitly integrate equation (5.4), which gives us
the following answer.

Let $v_{\pm}$ be the basis of $V^1$, such that $hv_{\pm}=\pm v_{\pm},\
ev_-=v_+,\ fv_+=v_-,\ ev_+=fv_-=0$.
 Let $\eta(q)$ be the Dedekind function:
$$
\eta(q)=q^{1/24}\prod_{m=1}^{\infty}(1-q^m).\tag 5.6
$$
Then
$$
T_{\pm}(q)=\eta(q^2)^{\frac{3}{4\kappa}}v_{\pm}.
\tag 5.7
$$
Since the Dedekind function is a modular form of weight $1/2$,
the function (5.7) is a
modular function of weight $3/8\kappa$, which is by no means a surprise in
view of formula (4.16).

{\bf Example 2. } Let $\Phi(z):M_{-\lambda,k}\to
\hat M_{\lambda,k}\otimes V^2_C(z)$ be an intertwining operator,
where $V^2$ is the three-dimensional irreducible representation of
${\frak sl}_2$. Then we must have $\lambda=0 \text{ or }\pm 1$.
Denote the corresponding operators by $\Phi^0,\Phi^{\pm}$, and
introduce the notation: $T_0(q)=\text{Tr}(\Phi^0Bq^{-\partial})$,
$T_{\pm}(q)=\text{Tr}(\Phi^{\pm}Bq^{-\partial})$.
These traces are computed similarly to Example 1. The matrix $L(1,q)$
is now no longer scalar, but it is a diagonal matrix, so
equation (5.4) is still easy to integrate. Here is the answer:

Let $v_+,v_0,v_-$ be the basis of $V^2$, such that $hv_+=2v_+$,
$hv_-=-2v_-$, $hv_0=0$, $ev_-=fv_+=v_0$. Then:
$$
\gather
T_0(q)=\eta(q^2)^{\frac{4}{\kappa}}\eta(q^4)^{-\frac{2}{\kappa}}v_0,\\
T_{\pm}(q)=\frac{1}{2}\eta(q^4)^{\frac{1}{\kappa}}\eta(q^2)^{\frac{1}{\kappa}}
\biggl[\biggl(\frac{\eta(q)}{\eta(-q)}\biggr)^{\frac{1}{\kappa}}(v_++v_-)
\pm\biggl(\frac{\eta(-q)}{\eta(q)}\biggr)^{\frac{1}{\kappa}}(v_+-v_-)\biggr].
\tag
5.8
\endgather
$$

Let us now calculate a simplest 2-point trace (a part of this computation
is due to A.Kirillov Jr.).

{\bf Example 3.} Consider intertwining operators:
$\Phi^{\pm}(z):M_{\lambda,k}\to \hat M_{\lambda\pm 1,k}\otimes
V^1_C(z)$. We can combine four traces out of these operators:
$$
\gather
T_{\pm\pm}(z|q)=\text{Tr}(\Phi^{\pm}(z_1)\Phi^{\pm}(z_2)Bq^{-\partial}),
z=\frac{z_1}{z_2}.
\tag
5.9\endgather
$$
It is possible to calculate these traces explicitly using the elliptic
KZ equations.

We have proved that the traces (5.9) satisfy the equation
$$
\kappa\frac{\partial F}{\partial z}=\rho(z|q)F\tag 5.10
$$

In the case under consideration, the traces take values in the four
dimensional space $V^1_C\otimes V^1_C$, and this
equation can be explicitly solved. Indeed, let us seek the solution in
the form
$$
g(z|q)=g_{++}(z|q)v_+\otimes v_+ +g_{+-}(z|q)v_+\otimes v_- +
g_{-+}(z|q)v_-\otimes v_+ + g_{--}(z|q))v_-\otimes v_-.\tag 5.11
$$

Consider the functions
$$
\gather
h_1(z|q)=\frac{1}{2}(g_{++}(z|q)+g_{--}(z,q)),\\
h_2(z|q)=\frac{1}{2}(g_{++}(z|q)-g_{--}(z,q)),\\
h_3(z|q)=\frac{1}{2}(g_{+-}(z|q)+g_{-+}(z,q)),\\
h_4(z|q)=\frac{1}{2}(g_{+-}(z|q)-g_{-+}(z,q)),
\tag 5.12
\endgather
$$
We have
$$
g(z|q)=h_1(z|q)(v_{++}-v_{--})+h_2(z|q)(v_{++}+v_{--})+h_3(z|q)(v_{+-}-v_{-+})
+h_4(z|q)(v_{+-}+v_{-+}).\tag
5.13
$$

Equation (5.10) yields a separate first order linear differential
equation  for each of the functions $h_1,h_2,h_3,h_4$:
$$
\gather
\kappa\frac{\partial h_1}{\partial z}=\frac{1}{2}(-a+c)h_1,\\
\kappa\frac{\partial h_2}{\partial z}=\frac{1}{2}(a+c)h_2,\\
\kappa\frac{\partial h_3}{\partial z}=\frac{1}{2}(-b-c)h_3,\\
\kappa\frac{\partial h_4}{\partial z}=\frac{1}{2}(b-c)h_4.
\tag 5.14\endgather
$$

These equations are easily solved. To write down the solutions,
it is convenient to use the function
$$
\gather
E(z|q)=-\frac{1}{2}\wp^{\prime}\biggl(\frac{\log z}{2\pi
\text{i}}|2\tau\biggr),\\
-\frac{1}{2}\wp^{\prime}(y|2\tau)=\sum_{m,p\in\Bbb Z}(y-m-2p\tau)^{-3}
\tag 5.15\endgather
$$
In terms of this function,
the solutions of (5.14) can be written in the form:
$$
\gather
h_1(z|q)=C_1(q)E(-z|q)^{\frac{1}{4\kappa}},\\
h_2(z|q)=C_2(q)E(-qz|q)^{\frac{1}{4\kappa}},\\
h_3(z|q)=C_3(q)E(z|q)^{\frac{1}{4\kappa}},\\
h_4(z|q)=C_4(q)E(qz|q)^{\frac{1}{4\kappa}},\\
\tag
5.16\endgather
$$

The coefficients $C_j(q)$ are easily found from equation (3.24):
$$
\gather
C_1(q)=C_1\eta(q^4)^{\frac{2}{\kappa}}\eta(q^2)^{-\frac{1}{\kappa}}\biggl
(\frac{\eta(-q)}{\eta(q)}\biggr)^{\frac{1}{\kappa}},\\
C_2(q)=C_2\eta(q^4)^{\frac{2}{\kappa}}\eta(q^2)^{-\frac{1}{\kappa}}\biggl
(\frac{\eta(q)}{\eta(-q)}\biggr)^{\frac{1}{\kappa}},\\
C_3(q)=C_3,\\
C_4(q)=C_4\eta(q^2)^{\frac{4}{\kappa}}\eta(q^4)^{-\frac{2}{\kappa}},\\
\tag 5.17\endgather
$$

It remains to say what values of constants $C_j$ correspond to the
traces (5.9). This information is given below:
$$
\gather
T_{++}:\ C_1=C_2=1,\ C_3=C_4=0;\\
T_{--}:\ C_1=-1,\ C_2=1,\ C_3=C_4=0;\\
T_{-+}:\ C_1=C_2=0,\ C_3=-1,\ C_4=1;\\
T_{+-}:\ C_1=C_2=0,\ C_3=C_4=1.\tag 5.18\endgather
$$

In general, solutions of the elliptic KZ equations cannot be expressed
in terms of classical
elliptic and modular functions. Their components are more complicated special
functions associated with an elliptic curve.

{\bf Example 4. } Consider the elliptic KZ equations with coefficients
in $V_C^1\otimes V_C^3$. Let $v_1, v_{-1}$ be the basis of $V_C^1$
introduced in Example 1 (earlier we used the notation $v_+,v_-$ for this
basis), and let
$w_{-3},w_{-1},w_1,w_3$ be a basis of $V_C^3$ in which
$fw_3=w_1,\ fw_1=2w_{-1},\ fw_{-1}=3w_{-3},\ fw_{-3}=0$,
$ew_{-3}=w_{-1},\ ew_{-1}=2w_{1},\ ew_{1}=3w_{3},\ ew_3=0$
. Let us look for
solutions of the elliptic KZ of the form:
$$
\gather
F(z|q)=h_+(z|q)(v_1\otimes w_1+v_{-1}\otimes w_{-1})+
h_-(z|q)(v_{-1}\otimes w_{-1}-v_1\otimes w_1)+\\
f_+(z|q)(v_1\otimes w_{-3}+v_{-1}\otimes w_3)+
f_-(z|q)(v_1\otimes w_{-3}-v_{-1}\otimes w_3).\tag 5.19\endgather
$$
The functions $h_{\pm}$ and $f_{\pm}$ can be found from the following
$2\times 2$ linear systems of differential equations:
$$
\gather
\kappa\frac{\partial h_+}{\partial
z}=(\frac{1}{2}c+ a)h_++\frac{1}{2}bf_+,\\
\kappa\frac{\partial f_+}{\partial
z}=\frac{3}{2}bh_+-\frac{3}{2}cf_+
;\tag 5.20
\endgather
$$
$$
\gather
\kappa\frac{\partial h_-}{\partial
z}=(\frac{1}{2}c- a)h_-+\frac{1}{2}bf_-,\\
\kappa\frac{\partial f_-}{\partial
z}=\frac{3}{2}bh_--\frac{3}{2}cf_-
.\tag 5.21
\endgather
$$

Let us reduce the systems (5.20) and (5.21) to second order
differential equations. It will be convenient to use the notation
$\phi_{\pm}=\frac{1}{2}c\pm a$. Differentiating the first equation in
(5.20) and (5.21), we get
$$
\kappa h_{\pm}^{\prime\prime}=\phi_{\pm}^{\prime}h_{\pm}
+\phi_{\pm}h_{\pm}^{\prime}+
\frac{1}{2}bf_{\pm}^{\prime}+\frac{1}{2}b^{\prime}f_{\pm}
. \tag 5.22
$$
Substituting the second equation in (5.20), (5.21) into (5.22), we obtain
$$
\kappa h_{\pm}^{\prime\prime}=\phi_{\pm}^{\prime}h_{\pm}
+\phi_{\pm}h_{\pm}^{\prime}+\frac{1}{2\kappa}b\biggl(\frac{3}{2}bh_{\pm}-
\frac{3}{2}cf_{\pm}\biggr)+\frac{1}{2}b^{\prime}f_{\pm}
.\tag 5.23
$$
Substituting the relation $f_{\pm}=\frac{2}{b}(\kappa
h_{\pm}^{\prime}-\phi h_{\pm})$ into (5.23), and taking into account
the
fact that
$-b^{\prime}/b=2c$,
we deduce the second order equation
for $h$:
$$
h_{\pm}^{\prime\prime}+\biggl((2+\kappa^{-1})c\mp\kappa^{-1}a\biggr)
h_{\pm}^{\prime}-
\kappa^{-1}\biggl(\phi_{\pm}^{\prime}+\frac{3}{4\kappa}b^2+\biggl
(\frac{3}{2\kappa}+2\biggr)c\phi_{\pm}\biggr)h_{\pm}=0.\tag 5.24
$$

Equation (5.24) is a second order equation on the elliptic curve
$\Bbb C/<1,2\tau>$
with four regular singularities at the points of order 2. This
equation is obviously invariant under the transformation
$z\to -z$. Therefore,
if we make a change of variable $w=\wp(z|2\tau)$ (here
$\wp(z|2\tau)=-\zeta^{\prime}(z|2\tau)$ is the Weierstrass
$\wp$-function), we will obtain
a second order equation on $\Bbb CP^1$ with rational coefficients and
four regular singularities, at the points $E_1=\wp(1/2|2\tau)$,
$E_2=\wp(\tau|2\tau)$, $E_3=\wp(\tau+1/2|2\tau)$, and $\infty$.
By a linear transformation of the independent variable the singular
points can be placed at $0$, $1$, $\lambda$, and $\infty$, where
$\lambda=\frac{E_3-E_1}{E_2-E_1}$.

A general second order equation with four regular singularities
was first studied by K.Heun\cite{He} in 1889, and it is now called
Heun's equation. The general form of Heun's equation is
$$
u^{\prime\prime}+\biggl(\frac{\gamma}{x}+\frac{\delta}{x-1}+\frac{\epsilon}
{x-\lambda}\biggr)u^{\prime}+\frac{\alpha\beta
x-p}{x(x-1)(x-\lambda)}u=0,\tag 5.25
$$
where
$$
\alpha+\beta-\gamma-\delta-\epsilon+1=0.\tag 5.26
$$
Its solutions (which are called Heun's functions)
were studied in much detail in the middle of the
20-th century\cite{Er}.

A Heun's equation is characterized by a $P$-symbol which contains the
information about the eigenvalues of monodromy matrices at singular
points (see \cite{Er}). The $P$-symbol is a table whose columns
correspond to the singular points of the equation. In each column
there are three numbers -- the coordinate of the singular point and
the two roots of the characteristic equation at this point.
The $P$-symbol of (5.25) is
$$
P\bigg\lbrace\matrix 0&1&\lambda&\infty&\ \\ 0&0&0&
\alpha &z\\1-\gamma & 1-\delta &
1-\epsilon &\beta &\ \endmatrix\bigg\rbrace
.\tag 5.28
$$
 Of course, the $P$-symbol does not determine the
equation uniquely since it contains no information about the constant
$p$.

 In the special case of equation (5.24), we can easily write down
the $P$-symbol. The $P$-symbol of the transformed
equation (5.25) for the plus
sign is
$$
P\bigg\lbrace\matrix 0&1&\lambda&\infty&\ \\ 0&0&0&
-\frac{3}{2\kappa}+\frac{3}{4}&z\\ \frac{1}{2}&\frac{1}{\kappa}&
\frac{1}{\kappa}&-\frac{1}{2\kappa}+\frac{3}{4}&\ \endmatrix\bigg\rbrace
.\tag 5.29
$$
For the minus sign the columns under $0$ and $\lambda$
 in (5.29) have to be interchanged.

It is easy to see that equations (5.20) and (5.21) (and hence (5.24))
cannot be solved
in elementary and elliptic functions. To verify this fact it is enough to look
at the limit $q\to 0$.In this limit we are supposed to get the
solution of the KZ equations in the space $V^1\otimes V^3$. This
solution is known to be given by a hypergeometric function. The
combination of parameters in this specific case is such that this
hypergeometric function does not express through elementary functions.
This implies that solutions of systems (5.20), (5.21)
cannot be expressed in terms of elliptic functions.
\vskip .1in

\heading
{\bf 6. Monodromy of the elliptic KZ equations}
\endheading

In this section we will study the monodromy of the elliptic KZ
equations with respect to the lattice of periods, and compute
the monodromy matrices. Although it is difficult to compute the
solutions, the calculation of monodromy is fairly straightforward.

Let us first describe how to interchange the order of intertwining
operators.

Let $\Phi^{w,\lambda,\nu}(z):M_{\lambda,k}\to M_{\nu,k}\hat\otimes
z^{\Delta}V_C(z)$ be the intertwining operator such that
\linebreak $<v_{\nu}^*,\Phi^{w,\lambda,\nu}(z)v_{\lambda}>=w$, $w\in
V^{\lambda-\nu}$. Suppose that $z_1,z_2$ are nonzero complex numbers,
and we have a product
$\Phi^{w_1,\lambda_1,\lambda_0}(z_1)\Phi^{w_2,\lambda_2,\lambda_1}(z_2)
:M_{\lambda_2,k}\to \hat M_{\lambda_0,k}\otimes V_1\otimes V_2$ where
$V_1$ and $V_2$ are finite dimensional representations of $\frak g$.
The question is: can this product be expressed in terms of products of
the form $\Phi(z_2)\Phi(z_1)$? Of course, we can only talk about such
an expression after analytic continuation, since the former is defined
for $|z_1|>|z_2|$, and the latter for $|z_1|<|z_2|$. However, if we
apply analytic continuation, the answer to the question is positive,
and given by the following theorem.

\proclaim{Theorem 6.1}(see \cite{FR},\cite{TK}) Let $x_{i\nu}$ be a basis of
$V_1^{\nu-\lambda_0}$, and let $y_{i\nu}$ be a basis of
$V_2^{\lambda_2-\nu}$. Then
$$
\Phi^{x_{r\lambda_1},\lambda_1,\lambda_0}(z_1)
\Phi^{y_{s\lambda_1},\lambda_2,\lambda_1}(z_2)=
A^{\pm}\sum_{\nu,i,j}\check
R^{\pm}_{ijrs\lambda_1\nu}(\lambda_2,\lambda_0)^{V_1V_2}
\sigma\Phi^{y_{i\nu},\nu,\lambda_0}(z_2)
\Phi^{x_{j\nu},\lambda_2,\nu}(z_1),\tag 6.1
$$
where $A^{\pm}$ is the analytic continuation along a path in which
$z_1$ passes $z_2$ from the right (for plus) and from the left (for
minus), respectively, $\check R^{\pm}(\lambda,\mu)^{V_1V_2}$ is a matrix,
and $\sigma$ is the permutation: $V_1\otimes
V_2\to V_2\otimes V_1$.
\endproclaim

Clearly, the matrix $\check R^{\pm}(\lambda,\mu)^{V_1V_2}$
represents a linear operator\linebreak
$(V_1\otimes V_2)^{\lambda-\mu}\to(V_2\otimes V_1)^{\lambda-\mu}$. Therefore,
if we define
$$
\check R^{\pm}(\lambda)^{V_1V_2}
=\oplus_{\mu}\check R^{\pm}(\lambda,\mu)^{V_1V_2},\tag 6.2
$$
then $\check R^{\pm}(\lambda)^{V_1V_2}$ will correspond to an operator:
 $V_1\otimes V_2\to V_2\otimes V_1$. This operator has been
computed\cite{Koh},\cite{Dr},\cite{TK},\cite{SV}, and it turned out to
be
proportional to the product of the
quantum $R$-matrix of the quantum group $U_{\hat q}(\frak g)$ and the
permutation $\sigma$ (the order of this product depends on the sign,
plus or minus), where $\hat q=e^{2\pi\text{i}/N\kappa}$.
Hereafter we will assume that the matrix $\check R^{\pm}(\lambda)^{V_1V_2}$ is
known.

Let us now compute the monodromy of solutions of the elliptic KZ equations.
Again, we will need to assume that $V_1,...,V_n$ are
finite dimensional representations of $GL_N$.

The elliptic KZ system is a local system with singularities, $\Cal L$,
 on the
space $E^n$, where $E$ is the elliptic curve $\Bbb C^*/\Gamma$,
and $\Gamma$ is the multiplicative subgroup in $\Bbb C^*$ generated by
$q^N$. The fiber of this local system is $V_1\otimes\dots\otimes V_n$.
Such an interpretation, however, is not very convenient for
computation of monodromy since $\Cal L$ has too many
singularities: they occur whenever $z_i/z_j=\varepsilon^mq^p$.
It would be more natural to regard the elliptic KZ system as
a local system on the $n$-th power of
a smaller elliptic curve $\hat E=\Bbb C^*/\hat\Gamma$, where
$\hat\Gamma$ is generated by $q$ and $\varepsilon$.
In this case, the singularities would occur only on the loci
$z_i=z_j$. But unfortunately, the elliptic KZ system is not a local system
with singularities on
$\hat E^n$: its right hand side is not $q,\varepsilon$-periodic.
Therefore we would like to produce a local system on $\hat E^n$ starting with
$\Cal L$. For this purpose we will use the fact that $\Cal L$ has a
finite group of symmetries.

Recall the notation of Section 4: $H_N$ denotes the Heisenberg group
of order $N^3$. Let $H_N^n$ denote the $n$-th Cartesian power of
$H_N$. Since the group $H_N$ is naturally embedded in $GL_N$
(it is generated by $B$ and $C$), we have a natural representation of $H_N$ in
$V_i$ and hence a representation of $H_N^n$
in $V_1\otimes\dots \otimes V_n$. On the other hand, the group
$H_N$ operates on $E$: $Bz=qz,\ Cz=\varepsilon^{-1} z$, so the group
$H_N^n$ naturally operates on $E^n$. These two actions can be combined
into an action of $H_N^n$ in the trivial bundle over $E_N$ with fiber
$V_1\otimes\dots\otimes V_n$. This action has the following property.

\proclaim{Lemma} The group $H_N^n$ preserves the local system $\Cal
L$.
\endproclaim

\demo{Proof} The lemma follows from the definition of the elliptic
$r$-matrix (formula (3.13)).
\enddemo

Now we can create a new ``local system``
$\Cal S=\Cal L/H_N^n$. The fiber of this local system is no longer the
space $V_1\otimes \dots\otimes V_n$ but rather the quotient of this
space by the action of the center of $H_N^n$ (which is, of course, not
a vector space).
In this section we will describe the monodromy of this local system.
This monodromy will be a linear representation of a suitable central
extension of the fundamental group of $\hat
E^n\backslash\text{\lbrace diagonals\rbrace}$, and it
obviously contains all the information about the
monodromy of the elliptic KZ equations.

As before, we will use the fundamental solution $\Cal
F(z_1,...,z_n|q)$
defined by (4.18).
This solution takes values in the space
$\text{End}(V_1\otimes\dots\otimes V_n)$.
The fundamental solution has a defining property:
if $u$ is a vector in $V_1\otimes\dots\otimes V_n$ then
$\Cal Fu$ is the solution of the system (3.14), (3.24) which
tends to $u$ as $q\to 0$ and $z_i/z_{i+1}\to \infty$.

Below we study the monodromy of the fundamental solution.
This is equivalent to studying the monodromy of the local system $\Cal S$.

We will use the representation of $\hat E$ as a parallelogram on the
complex plane: $\frac{\log z}{2\pi
i}=\frac{x}{N}+y\tau$, $0\le x,y<1$, and write $z=(x,y)$.
We choose a base point
$(z_1^0,...,z_n^0)=(x_1^0,y_1^0,...,x_n^0,y_n^0)$ on $\hat E_n$ such that
$1> x_1^0>\dots>x_n^0>0$,
$1> y_1^0>\dots>y_n^0>0$.

It follows from the definition
of $\Cal F(z_1,...,z_n|q)$ that it can be represented in the form
$z_1^{D_1}z_2^{D_2}\dots z_n^{D_n}\Cal F_0(z_1,...,z_n|q)$
where $D_1,\dots,D_n$ are operators in $V_1\otimes\dots\otimes V_n$
defined as follows. Let vectors $v_j\in V_j$ satisfy the condition
$hv_j=\chi_j(h)v_j$, $h\in{\frak h}$, $\chi_j\in {\frak h}^*$.
Let
$$\lambda_j=(\beta-1)^{-1}(\sum_{i=1}^n\chi_i)+\sum_{i=1}^j\chi_i,\
0\le j\le n.\tag
6.3
$$
Then
$$
D_j(v_1\otimes\dots\otimes
v_n)=\frac{<\lambda_j,\lambda_j>-<\lambda_{j-1},\lambda_{j-1}>}{2\kappa}\cdot
v_1\otimes\dots\otimes v_n.\tag 6.4
$$
(from this definition of $D_j$ it immediately follows that
$[D_i,D_j]=0$ for any $i,j$).

This observation helps us find the monodromy of the function $\Cal F$
around the $\varepsilon$-cycles on $\hat E^n$.
If $z_j$ is rotated around the origin
anticlockwise, through the angle $2\pi\text{i}l/N$,
with the rest of the variables fixed,
the function $\Cal F$ multiplies by the matrix
$e^{2\pi \text{i}lD_j/N}$ from the right, and undergoes a conjugation by
$C_j$-the action of $C$ in $V_j$.
 We denote these monodromy operators by $E_j$:
$$
E_j(\Cal F)=C_j^{-1}\Cal F C_je^{2\pi \text{i}lD_j/N}.\tag
6.4
$$

The monodromy of the fundamental solution in the neighborhood of the
locus $z_j=z_{j+1}$ can be found with the help of Theorem 6.1.
Using this theorem and the representation of solutions as traces of
products of intertwiners, we immediately find that the monodromy
of the function $\Cal F$,
around the locus $z_j=z_{j+1}$
(in the anticlockwise direction) is
 $$b_{j,j+1}(\Cal F)=\Cal FS_{j,j+1},\tag 6.5$$
where
$$S_{j,j+1}=\check R^-_j(V_1,\dots,V_{j+1},V_j,\dots,V_n)^{-1}
\check R^+_j(V_1,\dots
V_j,V_{j+1},\dots,V_n),$$
and the linear operators
$$
\check R^{\pm}_j(V_1,...,V_n): V_1\otimes\dots \otimes V_j\otimes
V_{j+1}\otimes\dots\otimes V_n\to V_1\otimes\dots \otimes V_{j+1}\otimes
V_{j}\otimes\dots\otimes V_n
$$
are defined as follows: if $v_i\in V_i$, $1\le i\le n$, and $hv_i=
\chi_i(h)v_i$, $\chi_i\in {\frak h}^*$, $h\in {\frak h}$, then
$$
\check R^{\pm}_j(V_1,...,V_n)(v_1\otimes \dots\otimes v_j\otimes
v_{j+1}\otimes\dots\otimes v_n)=v_1\otimes \dots\otimes
\check R^{\pm}(\lambda_{j+1})^{V_jV_{j+1}}(v_j\otimes
v_{j+1})\otimes\dots\otimes v_n,
$$
where $\lambda_j$ are defined by (6.3).

 This monodromy is obviously the same
as for the usual (trigonometric) KZ equations, whose solutions are
given by matrix elements of intertwiners rather than traces (see
section 2) -- a known fact which was first proved by
I.Cherednik\cite{Ch2}.

Let us now find the monodromy of the function $\Cal F$
around the $q$-cycles. Consider the cycle on $\hat E^n$ in which $z_j$ passes
$z_{j+1},...,z_n$ from the right, hits the circle $|z|=|q|$
(which is identified with the circle $|z|=1$ through the map $z\mapsto
qz$), jumps over to the circle $|z|=1$, and then passes
$z_1,...,z_{j-1}$ from the left, returning to its initial position
(this corresponds to the path $y_j(t)=y_j^0-t \text{ mod }1$, with the
rest of $x_i$ and $y_i$ remaining unchanged).
Using Theorem 6.1 and the expression of $\Cal F$ in terms of traces,
we find the monodromy matrices $Q_j$ for the described $q$-cycles.
Indeed, we have to interchange $\Phi(z_j)$ in the trace expression
with $\Phi(z_{j+1})$,...,$\Phi(z_n)$, then with $q^{-\partial}$ and
$B$, and then with $\Phi(z_1)$,...,$\Phi(z_{j-1})$. This results in
an expression of $Q_j$ as a product of $R$-matrices. To write this
expression down, introduce
the following notation: if $s\in S_n$ is a permutation of $n$ items
then set $\check R^{\pm}_j(s)=\check
R^{\pm}_j(V_{s(1)},\dots,V_{s(n)}$.
Let $t_j$ be the elementary transpositions $(j,j+1)$, and let
$s_{jm}= t_{m-1}\dots t_{j+1}t_j$, $j<m\le n$, $s_{jm}=t_m\dots
t_{j-2}t_{j-1}$, $1\le m<j$ (we make a convention that for two
permutations $\sigma_1,\sigma_2$
$\sigma_1\sigma_2(j)=\sigma_1(\sigma_2(j))$, $1\le j\le n$, i.e. the
factors in a product of permutations are applied from right to left).
 Then the operators $Q_j$ are expressed as
follows:
$$
Q_j(\Cal F)=B_j\Cal F \check R_j^+(\text{Id})\check
R_{j+1}^+(s_{jj+1})
\dots  \check R_{n-1}^+(s_{jn-1})B_j^{-1} \check R_{1}^-(s_{j1})\dots
\check R_{j-1}^-(s_{jj-1}),
\tag 6.6
$$
where
$B_j$ denotes the action of $B$ in $V_j$.

{\bf Remarks. } 1. Expressions similar to (6.6) (ordered products of
$R$-matrices) occur in the theory of correlation functions for quantum
affine algebras. Such correlation functions satisfy a quantum analogue
of the Knizhnik-Zamolodchikov equations - a system of difference
equations discovered by Frenkel and Reshetikhin\cite{FR}. The structure of the
right hand side of this system resembles (6.6). It is not clear if
it is merely a coincidence or not.

2. It is seen from the definitions of $E_j$ and $Q_j$ that
$[E_i,E_j]=[Q_i,Q_j]=0$ for any $i,j$.
\vskip .1in

As we have already remarked,
the monodromy of the local system $\Cal S$
defines a representation of a central extension
of the fundamental group of the complement
of the diagonals $z_i=z_j$ in $\hat E_n$ --
the pure braid group of the torus. To describe this representation in
more detail, let us assume that  $V_i=V$ for all $i$, where $V$ is
some finite-dimensional $GL_N$-module. This does not cause any loss
of generality since we can always set $V=V_1\oplus\dots\oplus V_n$ to
include the previously considered case. But now the elliptic KZ system
(and the local system $\Cal S$)
has an additional symmetry -- the symmetry under the simultaneous
permutation of the variables $z_i$ and the factors in the product
$V\otimes V\otimes\dots\otimes V$. Therefore, the monodromy
representation can in fact be regarded as a representation of a
certain (in general, not central) extension of the
full braid group of the torus.

The braid group of the torus, $BT_n$, is generated by the elements
$T_i$, $1\le i\le n-1$, $X_1,Y_1$, satisfying the
defining relations
$$
\gather
T_iT_{i+1}T_i=T_{i+1}T_iT_{i+1};\ T_iT_j=T_jT_i,\ j>i+1;\\
(T_1X_1)^2=(X_1T_1)^2;\ (T_1Y_1^{-1})^2=(Y_1^{-1}T_1)^2;\\
T_jX_1=X_1T_j,\  T_jY_1=Y_1T_j,\ j>1;\\
X_2Y_1^{-1}X_2^{-1}Y_1=T_1^2,\ \text{where } X_{i+1}=T_iX_iT_i;\\
X_0Y_1=Y_1X_0,\ \text{where } X_0=X_1X_2\dots X_N
\tag 6.7\endgather
$$

It is also convenient to define the elements
$Y_{j+1}=T_j^{-1}Y_jT_j^{-1}$
and $Y_0=Y_1Y_2\dots Y_n$.

To picture the braid group of the torus geometrically,
 one should imagine $n$ ``beetles'' crawling on the
surface of the torus $\Bbb C/(\Bbb Z+\tau\Bbb Z)$
starting from some fixed positions $z_1,...,z_n$
($z_j=N^{-1}x_j+\tau y_j$, $x_1>x_2>...>x_n$, $y_1>y_2>...>y_n$) so
that at no time two beetles can be at the same point and after some
period of time (say 1) the ``beetles'' return to their original positions,
possibly with some permutation. Then the beetles will trace out some
collection of curves in $\text{Torus}\times [0,1]$ -- a braid diagram.
Such diagrams can be composed by attaching the bottom of one of them to
the top of the other. Under this composition law, braid diagrams form a
group -- the braid group of the torus $BT_n$. The element
 $T_j$ corresponds to the intertwining of the $j$-th and
$j+1$-th braids (the $j$-th and $j+1$-th ``beetles'' switch, the $j$-th
``beetle'' passing the $j+1$-th one from the right),
and the elements $X_j$ and $Y_j$  arise when the $j$-th ``beetle''
crawls around the $x$-cycle and $y$-cycle of the torus, respectively,
in the negative direction of the $x$-axis (respectively, $y$-axis),
with the rest of the ``beetles'' unmoved.

Now we are in a position to formulate the result about the monodromy
of the local system $\Cal S$.

\proclaim{Theorem 6.2}
The monodromy representation of an extension of
$BT_n$ associated to the local system $\Cal S$
is defined as follows:
$$
\gather \ X_j\mapsto e^{-2\pi \text{i}lD_j/N}C_j^{-1},\\
Y_j\mapsto \check R_j^+\check
R_{j+1}^+
\dots\check  R_{n-1}^+B_j^{-1}\check R_{1}^-\dots
\check R_{j-1}^- \\ T_j\mapsto (\check R_j^+)^{-1}.\tag 6.8
\endgather
$$
\endproclaim

{\bf Remark. } Since all the spaces $V_j$ are the same, we have dropped
the permutations labeling the $R$-matrices.

\demo{Proof} The theorem follows from Theorem 6.1 and formulas
(6.4)-(6.6).
\enddemo

The extension of $BT_n$ involved in Theorem 6.2 is very easy
to describe. Consider the quotient of the group $BT_n$
by the relations $T_i^2=1$. Then we will obtain the group
$S_n\ltimes (\Bbb Z^n\oplus \Bbb Z^n)$, where $S_n$ acts on both
copies of $\Bbb Z^n$ by permutations of components.
Let $H$ be the Heisenberg group -- the central extension of
$\Bbb Z\oplus \Bbb Z$ by $\Bbb Z$ by means of the 2-cocycle
$\omega(\bold x,\bold y)=x_1y_2-x_2y_1$. Then we can construct the
group $S_n\ltimes H^n$
($H^n$ is the $n$-th Cartesian power of $H$)
which is a rank $n$ abelian extension of $S_n\ltimes (\Bbb
Z^n\oplus\Bbb Z^n)$.
Let us denote the pullback of this abelian extension to $BT_n$
by $\widehat{BT_n}$.
The monodromy representation of $\Cal S$ is thus a representation of
$\widehat{BT}_n$ in $V\otimes\dots\otimes V$.

Now assume that $V$ is irreducible. Then instead of the group
$\widehat{BT}_n$ we will have to deal with a rank 1 central extension
$\widetilde{BT}_n$ of $BT_n$ which is constructed as follows. Take the group
$S_n\ltimes(\Bbb Z^n\oplus \Bbb Z^n)$, construct a rank $1$ central
extension of this group by adjoining a new central element $c$
satisfying the relations $X_iY_i=Y_iX_ic$, $X_iY_j=Y_jX_i$, $i\ne
j$, and then pull this extension back to $BT_n$. It follows then that
the monodromy representation of $\Cal S$ is a
projective representation of $BT_n$ which comes from a
linear representation of $\widetilde{BT}_n$.

The group $\widetilde{BT_n}$ is closely related to the double affine braid
group of type $A_{n-1}$ defined by I.Cherednik in his recent paper
\cite{Ch3} -- the group generated by the elements $T_i,X_i,Y_i$ and
a central element $\delta$ satisfying modified relations (6.7): the relation
$X_2Y_1^{-1}X_2^{-1}Y_1=T_1^2$ has to be replaced by
$X_2Y_1^{-1}X_2^{-1}Y_1=\delta T_1^2$.
Let us describe the connection between them.

Observe that the elliptic KZ system commutes with
the diagonal action of the Heisenberg group $H_N^{\text{diag}}$
generated by $X_0$
and $Y_0$ in
$V\otimes\dots\otimes V$. Decompose $V\otimes \dots\otimes V$ into a
sum of irreducible representations of $H_N^{\text{diag}}$:
$V\otimes\dots\otimes V=\bigoplus_{i}P_i\otimes W_i$, where $P_i$ are
distinct irreducible representations of $H_N$, and $W_i$ are
multiplicity spaces. Then each summand $P_i\otimes W_i$ is a
subrepresentation of $\widetilde{BT}_n$. Let
$\phi_i:\widetilde{BT}_n\to\text{End}(P_i\otimes W_i)$ be the
corresponding homomorphism.

Assume that $n$ and $N$ are coprime. Then we can define a
homomorphism $\xi:\widetilde{BT}_n\to H_N$
by $\xi(X_i)=C^{1/n}$, $\xi(Y_i)=B^{1/n}$,
$\xi(T_i)=1$. (Here $1/n$ is regarded as an element of
the ring $\Bbb Z/N\Bbb Z$.)
Composing this homomorphism with the diagonal action of $H_N$ in
$V\otimes\dots\otimes V$, we get a projective representation
$\psi_i(\cdot)$ of $BT_n$ in $\text{End}(P_i\otimes W_i)$.
 Notice that $\psi_i(g)=\phi_i(g)$ if $g\in H_N^{\text{diag}}$.
Let us write $\phi_i$ as $\phi_i(g)=
\psi_i(g)\cdot\psi_i(g)^{-1}\phi_i(g)$. Because the elliptic KZ system
commutes with $H_N^{\text{diag}}$, $\psi_i(g)$ commutes
with $\psi_i(g)^{-1}\phi_i(g)$, which implies that
$\psi_i(g)^{-1}\phi_i(g)=\text{Id}\otimes \chi_i(g)$, where
$\chi_i(\cdot)$ is some projective action of ${BT}_n$ in $W_i$.
We also have $\psi_i(g)=\tilde\psi_i(g)\otimes 1$, where
$\tilde\psi_i$ is an action of the group in $P_i$.
Therefore, we have $\phi_i=\tilde\psi_i\otimes\chi_i$. Therefore, we can
easily compute the 2-cocycle on $BT_n$ corresponding to $\chi_i$
as the difference of the 2-cocycles for $\phi_i$ and $\psi$. This
cocycle is the pullback from $S_n\ltimes (\Bbb Z^n\oplus\Bbb Z^n)$ of
the 2-cocycle given by:
$$
\gather
\omega\bigl((s_1,\bold x^1,\bold y^1),(s_2,\bold x^2,\bold y^2)\bigr)=
\sum_{i=1}^n(x_i^1y_i^2-x_i^2y_i^1)-\frac{1}{n}\sum_{i,j=1}^n(x_i^1y_j^2-
x_i^2y_j^1),\\
s_{1,2}\in S_n,\ \bold x^{1,2},\bold y^{1,2}\in \Bbb Z^n.\tag 6.9\endgather
$$
The extension of $BT_n$ by means of this cocycle is exactly the double
affine braid group. We denote this group by $\overline{BT}_n$.

The special case when $V$ is the $N$-dimensional vector representation
of $GL_N$ is especially interesting. In this case, we are
getting a representation of the {\it double affine Hecke algebra}
${\frak H}_n^{\hat q}$.
This algebra was recently defined by I.Cherednik \cite{Ch3}
as the quotient of the group algebra $\Bbb C[\overline{BT_n}]$ by the
relations $(T_j-\hat q)(T_j+\hat q^{-1})=0$. Indeed,
the matrix $\check R^+_j$ is diagonalizable and
has the eigenvalues $\hat q^{1-\frac{1}{2N}}$ and $-\hat
q^{-1-\frac{1}{2N}}$, (recall
$\hat q=e^{2\pi\text{i}/N\kappa}$). This statement easily follows from
the fact that this matrix is a monodromy matrix of
the elliptic KZ equations. Therefore, the matrix $\tilde
R^+_j=\hat q^{\frac{1}{2N}}\check R^+_j$ satisfies the equation
$$
(\tilde R^+_j-\hat q)(\tilde R^+_j+\hat q^{-1})=0.\tag
6.10
$$
Thus, the correspondence
$$
X_1\mapsto e^{-2\pi \text{i}lD_1/N}C_1^{-1},\ Y_1\mapsto
\tilde R_{1}^+ \tilde R_{2}^+
\dots  \tilde R_{n-1}^+B_1^{-1},\ T_j\mapsto \tilde R_j^+\tag 6.11
$$
defines a projective representation of $BT_n$ in the space
$V\otimes V\otimes\dots\otimes V$ which can be written as
$P\otimes W$, where $P$ is the $N$-dimensional irreducible
projective representation of $BT_n$ obtained by composing
the homomorphism $\xi$ defined above with the standard action
of $H_N$ in $\Bbb C^N$, and $W$ is a representation of the double
affine Hecke algebra ${\frak H}_n^{\hat q}$.

The element $\delta$ acts in the representation $W$ by multiplication by
${\hat q}^{-1/N}\varepsilon^{-1/n}$.

{\bf Remark. }
In fact, the monodromy
of the elliptic KZ system (3.14) extended by equation (3.24)
yields a representation of the semidirect product $SL_2(\Bbb
Z)\ltimes \widehat{BT}_n$ in the space
$\text{End}(V\otimes\dots\otimes V)$. As we
have already remarked in section 4, it is not clear
how to compute the modular part of this monodromy, i.e. the action of
$SL_2(\Bbb Z)$.
This computation, at least for one nontrivial example, is a very
interesting
and
challenging problem.
\vskip .1in

As a conclusion, let us note that the study of monodromy helps us to
find out for what special values of $\kappa$ the elliptic KZ equations
are integrable in elliptic functions.

\proclaim{Proposition 6.3} If $\kappa=1/MN$, where $M$ is an integer,
then the matrix elements of the fundamental solution $\Cal F(z_1,...,z_n|q)$
are finite products of rational powers of
theta functions of expessions $q^m\varepsilon^pz_i/z_j$, $0\le
m,n\le N-1$.
\endproclaim

\demo{Idea of
Proof} If $\kappa=1/NM$ then $\hat q=1$. Therefore, $\check R^+$
is a scalar matrix times the permutation of factors, so
$S_{j,j+1}=\alpha \times{Id}$, and $b_{j,j+1}(\Cal
F)=\alpha\Cal F$. Let $\alpha=e^{2\pi\text{i}s}$ ($s$ is
rational). Then we have
$$
\Cal F(z_1,...,z_n|q)=\prod_{i<j}\prod_{0\le m,p\le N-1}
\Theta(q^m\varepsilon^pz_i/z_j|q^N)^s\exp\biggl(\int\omega\biggr),
\tag 6.14
$$
where $\omega$ is a matrix-valued elliptic differential form on $E^n$ and
$$
\Theta(z|q)=\prod_{m\ge 0}(1-q^mz)(1-q^{m+1}z^{-1})(1-q^{m+1}).\tag
6.15
$$
\enddemo

{\bf Remark. }
A similar result holds for the conventional KZ equations (1):
if $k+h^{\vee}=1/M$, where $M$ is an integer then solutions of the KZ
equations are algebraic functions.

\end